\def\ls{\lower4pt\hbox{${\buildrel < \over \sim}$}}
\def\gs{\lower4pt\hbox{${\buildrel > \over \sim}$}}

\documentclass[12pt, preprint]{aastex}

\slugcomment{Accepted for Publication in {\it The Astrophysical Journal}}

\shorttitle{Multiwavelength Observations of PKS 0528+134}
\shortauthors{Palma et al.}

\begin{document}

\title{Multiwavelength Observations of the Gamma-Ray Blazar PKS~0528+134 in Quiescence}

\author{N. I. Palma\altaffilmark{1,2},
M. B\"ottcher\altaffilmark{1}, 
I. de la Calle\altaffilmark{3},
I. Agudo\altaffilmark{4,5}, 
M. Aller,\altaffilmark{6},
H. Aller,\altaffilmark{6},
U. Bach\altaffilmark{7},
E. Ben\'{\i}tez\altaffilmark{8},
C. S. Buemi\altaffilmark{9},
L. Escande\altaffilmark{10},
J. L. G\'{o}mez\altaffilmark{5}, 
M. A. Gurwell\altaffilmark{11}, 
J. Heidt\altaffilmark{12},
D. Hiriart\altaffilmark{13},
S. G. Jorstad\altaffilmark{4,15}, 
M. Joshi\altaffilmark{4}, 
A. L\"ahteenm\"aki\altaffilmark{14}, 
V. M. Larionov\altaffilmark{15,16},
P. Leto\altaffilmark{9},
Y. Li\altaffilmark{1},
J. M. L\'opez\altaffilmark{13},
B. Lott\altaffilmark{10},
G. Madejski\altaffilmark{17},
A. P. Marscher\altaffilmark{4}, 
D. A. Morozova\altaffilmark{15},
C. M. Raiteri\altaffilmark{18},
V. Roberts\altaffilmark{1},
M. Tornikoski\altaffilmark{14},
C. Trigilio\altaffilmark{9},
G. Umana\altaffilmark{9},
M. Villata\altaffilmark{18},
D. Wylezalek\altaffilmark{19}
}

\altaffiltext{1}{Astrophysical Institute, Department of Physics and Astronomy, 
Clippinger 339, Ohio University, Athens, OH 45701, USA}
\altaffiltext{2}{Facultad de Ciencias Espaciales, Universidad Nacional Autonoma de 
Honduras, Tegucigalpa M.D.C., Honduras C. A.}
\altaffiltext{3}{European Space Astronomy Center, P.O. Box 78, 
28691 Villanueva de la Ca$\tilde{\rm n}$ada, Madrid, Spain}
\altaffiltext{4}{Institute for Astrophysical Research, Boston
University, 725 Commonwealth Avenue, Boston, MA 02215, USA.; jorstad@bu.edu; 
marscher@bu.edu; iagudo@bu.edu}
\altaffiltext{5}{Instituto de Astrof\'{\i}sica de Andaluc\'{\i}a,
CSIC, Apartado 3004, 18080 Granada, Spain.; jlgomez@iaa.es}
\altaffiltext{6}{Department of Astronomy, University of Michigan, 
Ann Arbor, MI 48109-1042, USA}
\altaffiltext{7}{Max-Planck Institut f\"ur Radioastronomie, Auf dem H\"ugel 69,
D-53225 Bonn, Germany}
\altaffiltext{8}{Instituto de Astronom\'{\i}a, Universidad Nacional Aut\'onoma
de M\'exico, Apdo. Postal 70-264, CP 04510, M\'exico}
\altaffiltext{9}{INAF - Osservatorio Astrofisico di Catania, Italy} 
\altaffiltext{10}{Universit\'e Bordeaux 1, CNRS/IN2p3, Centre d'Etudes 
Nucl\'eaires de Bordeaux Gradignan, 33175 Gradignan, France}
\altaffiltext{11}{Harvard-Smithsonian Center for Astrophysics, Cambridge, MA} 
\altaffiltext{12}{ZAH, Landessternwarte Heidelberg, K\"onigstuhl,
D-69117 Heidelberg, Germany}
\altaffiltext{13}{Instituto de Astronom\'{\i}a, Universidad Nacional Aut\'onoma
de M\'exico, Apdo. Postal 877, CP 22800, Ensenada, B.C., M\'exico}
\altaffiltext{14}{Aalto University, Mets\"ahovi Radio Observatory,
Mets\"ahovintie 114, FIN-02540, Kylmala, Finland}  
\altaffiltext{15}{Astronomical Institute, St. Petersburg State University, 
Universitetsky pr. 28, Petrodvoretz, 198504 St. Petersburg, Russia}
\altaffiltext{16}{Isaac Newton Institute of Chile, St. Petersburg Branch, 
198504 St. Petersburg, Russia}
\altaffiltext{17}{Kavli Institute for Particle Astrophysics and Cosmology, 
Department of Physics and SLAC National Acelerator Laboratory, Stanford University, 
Stanford, CA 94305, USA}
\altaffiltext{18}{INAF, Osservatorio Astronomico di Torino, I-10025 Pino Torinese (TO), 
Italy}
\altaffiltext{19}{University of Cambridge, Department of Physics, Cavendish Laboratory, 
JJ Thomson Avenue, Cambridge, CB3 0HE, UK}

\begin{abstract} 

We present multiwavelength observations of the ultraluminous blazar-type radio 
loud quasar PKS~0528+134 in quiescence during the period July to December 2009.  
Four Target-of-Opportunity (ToO) observations with the {\it XMM-Newton} Satellite 
in the 0.2 -- 10 keV range were supplemented with optical observations at the MDM 
Observatory, radio and optical data from the GLAST-AGILE Support Program (GASP)
of the Whole Earth Blazar Telescope (WEBT) and the Very Long Baseline Array (VLBA), 
additional X-ray data from the {\it Rossi X-ray Timing Explorer} 
(RXTE; 2 -- 10~keV) and from {\it Suzaku} (0.5 -- 10~keV) as well as 
$\gamma$-ray data from the Fermi Large Area Telescope (LAT) in the 100~MeV 
-- 200 GeV range. In addition, publically available data from the
SMARTS blazar monitoring program and the University of Arizona / Steward
Observatory {\it Fermi} Support program were included in our analysis.

We found no evidence of significant flux or spectral variability 
in $\gamma$-rays and most radio bands. However, significant flux variability on 
a time scale of several hours was found in the optical regime, accompanied by a 
weak trend of spectral softening with increasing flux. We suggest that this 
might be the signature of a contribution of unbeamed emission, possibly from the
accretion disk, at the blue end of the optical spectrum.
The optical flux is weakly polarized with rapid variations of the degree and
direction of polarization, while the polarization of the 43~GHz radio core remains
steady, perpendicular to the jet direction. 
Optical spectropolarimetry of the object in the quiescent state suggests 
a trend of increasing degree of polarization with increasing wavelength,
providing additional evidence for an unpolarized emission component,
possibly thermal emission from the accretion disk, contributing towards 
the blue end of the optical spectrum. Over an extended period of several 
months, PKS~0528+134 shows moderate (amplitude $\lesssim 50$~\%)
flux variability in the X-rays and most radio 
frequencies on $\sim 1$ -- 2 week time scales. We constructed four spectral 
energy distributions (SEDs) corresponding to the times of the XMM-Newton 
observations. We find that even in the quiescent state, the bolometric 
luminosity of PKS~0528+134 is dominated by its $\gamma$-ray emission. 

A leptonic single-zone jet model produced acceptable fits to the 
SEDs with contributions to the high-energy emission from both synchrotron 
self-Compton radiation and Comptonization of direct accretion disk emission. 
Fit parameters close to equipartition between the energy densities of the
magnetic field and the relativistic electron population
were obtained. The moderate variability on
long time scales, compared to expected radiative cooling time scales, implies
the existence of on-going particle acceleration, while the observed optical
polarization variability seems to point towards a turbulent acceleration process. 
Turbulent particle acceleration at stationary features along the jet therefore 
appears to be a viable possibility for the quiescent state of PKS~0528+134. 
\end{abstract}


\section{\label{intro}Introduction}

Blazars (BL Lac objects and gamma-ray loud flat spectrum radio quasars [FSRQ]) 
are the most extreme type of active galactic nuclei (AGN). They were historically
defined through extreme flux variability throughout the electromagnetic spectrum, and 
sometimes strong and variable linear polarization at radio and optical wavelengths. 
In the 1990s, observations by {\it EGRET} on board the {\it Compton Gamma-Ray 
Observatory} revealed large $\gamma$-ray fluxes (often dominating the bolometric 
luminosity of the source) from many blazars.
The radio through optical emission from blazars is commonly interpreted as
synchrotron emission from ultrarelativistic electrons in a relativistic plasma 
jet that is closely aligned with our line of sight ($\theta_{obs} < 20^{\degr}$). 
This assertion
is supported by the superluminal motion that most blazars exhibit  
\citep[e.g.,][]{jorstad01,lister09,piner06} as well as by the observed luminosity and 
variability timescales observed in these objects. 
In extreme cases, variability time scales down to a few 
minutes have been found in the very high energy (VHE) $\gamma$-ray regime 
\citep[e.g.,][]{albert07,aharonian07}.

Two competing classes of models are currently being considered for the origin of
the high-energy (X-ray through $\gamma$-ray) emission from blazars. In leptonic
models, hadrons (primarily protons) in the jet (if present in substantial numbers
at all), are assumed not to be accelerated to ultrarelativistic energies.
They do not exceed
the threshold for photo-pion production processes on the low-frequency radiation 
field in the jet, and proton synchrotron radiation is assumed to be negligible.
Therefore, in leptonic models, the high-energy radiative output is dominated 
by Compton scattering of low-frequency photons off relativistic electrons. In
hadronic models, it is assumed that ultrarelativistic protons exist in sufficient
number. In such a scenario, the protons will dominate the radiative output via 
proton synchrotron radiation and synchrotron and Compton emission from secondary 
particles. 
Those are produced in photo-pion production and subsequent pion and 
muon decay and electromagnetic cascade processes.
For a recent review of blazar 
emission models, see, e.g., \cite{boettcher10}. 

The mechanism(s) of acceleration of particles to ultrarelativistic energies in
blazar jets are currently very poorly understood. Particle acceleration may be 
related to relativistic shocks in an unsteady flow \citep[e.g.,][]{mg85}, 
internal shocks resulting from the collision of relativistic plasma blobs 
ejected at different speeds \cite[e.g.,][]{spada01,mimica04,jb10,bd10}, 
re-collimation shocks \citep[e.g.,][]{bl09}, or relativistic shear layers
in radially stratified jets \citep[e.g.,][]{so02,rd04,rd06}, to name just 
a few plausible scenarios. Signatures that reveal the nature of particle 
acceleration in blazar jets may be found both in spectral and variability 
features. The nature of the acceleration mechanism is reflected in the 
shape of the produced particle spectra. Those, in turn, can be inferred from 
the shape of the non-thermal photon spectra, in particular in the synchrotron
part of the SED \citep[e.g.,][]{finke08}. The dynamics and light travel
time effects, in particular in shock acceleration scenarios, will leave 
distinct imprints in the observed variability features \citep[e.g.,][]{bd10}.

Observational studies of blazars have so far mostly focused on bright, flaring 
states of blazars. This is the consequence of observational constraints which 
make detailed measurements of spectral and variability features in X-rays and 
$\gamma$-rays difficult in low flux states. However, blazars are known to 
spend most of the time in their quiescent state which has so far received 
very little attention and is therefore very poorly understood. 
EGRET detected $\gamma$-ray blazars almost exclusively in flaring states, 
and the simultaneously operating X-ray telescopes 
(ROSAT, ASCA, RXTE) lacked the sensitivity to measure detailed X-ray spectral 
and variability properties of most blazars in their quiescent states. Therefore, 
even the question whether $\gamma$-ray emission persists at all in the quiescent 
states of blazars remained an open issue during the EGRET era. 

The observational situation has dramatically changed with the advent of the 
new generation of X-ray observatories, in particular {\it Chandra} and 
{\it XMM-Newton} as well as the launch of the {\it Fermi} Gamma-Ray Space 
Telescope in June 2008. The {\it Fermi} Large Area Telescope \citep[LAT,][]{atwood09} 
is continuously monitoring the entire sky every 3 hours in the energy range 
20~MeV -- 300~GeV with about an order of magnitude superior sensitivity 
compared to that of EGRET. It routinely detects $\gamma$-ray emission from 
known blazars even in their quiescent states. A detailed study of the quiescent 
state of blazars may elucidate whether the quiescent jet flow is smooth, 
exhibiting little or no variability, or the quiescent emission consists of 
the superposition of a rapid succession of ``mini-flares''. In particular, 
a featureless light curve in all bands might indicate the persistence of
particle acceleration mechanisms not related to impulsive (shock) events,
and might point towards shear-flow acceleration in radially stratified jets,
or standing features such as re-collimation shocks in the quiescent states
of blazars. 

This situation has motivated us to propose Target-of-Opportunity (ToO) observations
with {\it XMM-Newton} in AO-8, triggered by an extended quiescent state of a known
$\gamma$-ray bright blazar. We defined a quiescent state of a $\gamma$-ray blazar
by the object maintaining a $> 100$~MeV $\gamma$-ray flux lower than the lowest 
flux or upper limit ever determined by EGRET, over at least 2 weeks. The 
prominent high-redshift $\gamma$-ray bright FSRQ PKS~0528+134 fulfilled our 
pre-specified trigger criterion through its continued $\gamma$-ray quiescence 
for several months prior to September 2009. We therefore triggered our 
{\it XMM-Newton} ToO observations on PKS~0528+134.
The observations consisted of four observations on 2009 September 8, 10, 12, 
and 14. These were coordinated with ground-based radio and optical observations.
Simultaneous $\gamma$-ray observations were provided by {\it Fermi} LAT. 

In the following section, we give an overview of the known properties of
our target, PKS~0528+134. In \S \ref{observations}, observations and data 
reduction procedures are described. In \S \ref{variability}, we present the 
results of a flux and spectral variability analysis. The structure of the 
parsec scale jet of this source is presented in section \S \ref{scalejet}. 
Results of our modeling of four simultaneous SEDs obtained during our 
campaign are discussed in \S \ref{SED}. A discussion on the optical spectral 
variability and other relevant issues is presented in section \S \ref{discussion}. 
Finally, we summarize our results and draw conclusions in \S \ref{summary}.
Throughout this paper, we refer to a spectral index $\alpha$ as the energy
index such that $F_{\nu} \propto \nu^{-\alpha}$, corresponding to a photon
index $\Gamma_{\rm ph} = \alpha + 1$. We use a $\Lambda$CDM
cosmology with $\Omega_m = 0.3$, $\Omega_{\Lambda} = 0.7$, and $H_0 = 
70$~km~s$^{-1}$~Mpc$^{-1}$. In this cosmology, the luminosity distance
of PKS~0528+134 is $d_L = 16.2$~Gpc.

\section{\label{0528}The Quasar PKS~0528+134}

The compact FSRQ PKS 0528+134 is one of the most luminous and most distant 
$\gamma$-ray blazars known, with a redshift of $z = 2.07$  \citep{hunter93}. In 
the high-energy $\gamma$-ray band (above 100 MeV), this source was first detected 
by EGRET during the period 1991 April--June  \citep{mattox97}. Besides EGRET, 
this source was also detected by the other two instruments onboard of {\it CGRO}:
the Oriented Scintillation Spectrometer Experiment (OSSE) in the 0.05 -- 1.0 MeV 
band  \citep{mc95}, and the Imaging Compton Telescope (COMPTEL) ($\approx$ 0.75 
-- 10 MeV). During EGRET observations (from 1991 - 2000) PKS 0528+134 showed 
intense variability  \citep{dingus96} exhibiting strong flares in 1991, 1993, 
1995, and 1996. Its highest $\gamma$-ray flux was detected in 1993 March, when it 
reached $10^{14}$~JyHz \citep{muk96}, strongly dominating the bolometric luminosity
of the source. PKS~0528+134 is faint in the optical with a mean visual magnitude 
of $m_{v} = 19.5$ \citep{wall85}. This is a consequence of the high Galactic 
extinction in the direction of PKS 0528+134, estimated to be $2.782 < A_{v}< 5$ 
\citep{schlegel98,zhang94}. The reason for this high extinction is that 
PKS~0528+134 with Galactic coordinates $l = 191.37^{\degr}$, $b = -11.01^{\degr}$ 
($\alpha = 5^{h}30^{m}56^{s}.41, \delta = +13^{\degr}31'55.''15$ J2000), is 
located behind the 
translucent molecular cloud B30 \citep{liszt93,hogerheijde95}. Hence, there 
have been relatively few optical observations of this source compared to other 
bands.

In the radio regime, PKS~0528+134 is regularly monitored by several programs at 
different frequencies. The source shows pronounced radio flux-density variability 
on timescales of several months to a few years  
\citep{aller85,reich93,zhang94,stevens94,valtaoja95,pohl96,peng01,bach07}. 
A delay from high frequencies to low frequencies in radio bursts has been identified, 
and delays of a few months between $\gamma$-ray flares and the corresponding radio 
bursts have been found \citep{muk96}. Additionally, from VLBI observations 
performed in the 8.4~GHz band over a period of almost 8 years \cite{britzen99} found 
that PKS~0528+134 has a bent jet of length $\approx$ 5 -- 6 mas extending toward the 
northeast, in which rapid structural changes take place. Increasing activity in 
the radio and $\gamma$-ray bands is associated with morphological changes in the 
radio structure of the jet, and superluminal motion with $\beta_{\perp, \rm app} 
\lesssim 30$ is found in some of the jet components \citep{J05}.

Prior to the time period of EGRET, observations at X-ray energies were carried 
out by the Einstein Observatory in 1980, but no high confidence values for the 
X-ray flux and spectral index were found due to the low count statistics 
\citep{bregman85}. Starting in March 1991, and later in September 1992, 
X-ray observations in the energy range 0.07 -- 2.48 keV, performed with 
the ROSAT Position Sensitive Proportional Counter (PSPC)  were used 
to investigate the geometry and physical environment of PKS 0528+134 
\citep{zhang94,muk96}. The continuum emission of this source in  the medium -- hard 
X-ray band (0.4-10 keV) was first measured using observations with the Advanced 
Satellite for Cosmology and Astrophysics (ASCA) in 1994 and 1995 \citep{sambruna97}. 
Further X-ray observations of PKS~0528+134 have been carried out with the Rossi 
X-ray Timing Explorer (RXTE) during August and September 1996 and May 1999, as
well as with {\it BeppoSAX} in the 0.1 - 10 keV and 15-200 keV bands during 1997 
February and March as part of a multiwavelength campaign involving EGRET and 
ground based telescopes \citep{ghisellini99}. 

SEDs of PKS~0528+134 collected during six years of EGRET observations were
compiled and modeled with a one-zone leptonic jet model by \cite{muk99}.
It was found that during all EGRET detections of the source, the bolometric
luminosity was dominated by its $\gamma$-ray output. While, due to often
incomplete multiwavelength coverage, the modeling results of that paper
were subject to a large degree of freedom and uncertainty, the observed
epoch-to-epoch variability of PKS~0528+134 was found to be consistent with
a correlation between the $\gamma$-ray flux and the bulk Lorentz factor of
the emission region along the jet.

Given the extreme properties of PKS 0528+134, this blazar has been the target 
of many observations at different wavelengths. However, as for almost all 
blazars (see \S \ref{intro}), coordinated multiwavelength campaigns have 
targeted flaring states. The quiescent-state SEDs and spectral variability 
patterns of blazars in general, and of PKS 0528+134 in particular, are 
poorly understood. Given its extended $\gamma$-ray quiescence throughout
2009, as revealed by {\it Fermi} LAT minotoring\footnote{see \tt 
http://fermi.gsfc.nasa.gov/ssc/data/access/lat/msl\_lc/}, PKS~0528+134 
was therefore found to be an appealing target for our pre-approved 
{\it XMM-Newton} Cycle 8 ToO observations and multiwavelength campaign.

\section{\label{observations}Observations and Data Reduction} 

The blazar PKS 0528+134 was the target of intensive, simultaneous and 
quasi-simultaneous observations at optical (MDM, GASP), radio (GASP), 
X-ray ({\it XMM-Newton, RXTE, Suzaku}), and $\gamma$-ray  ({\it Fermi}~LAT)
frequencies during the period September 8 to 18, 2009. In 
addition, a more extended period of time, throughout July -- December 
2009, was covered by less intensive radio, optical, X-ray, and 
$\gamma$-ray monitoring for longer-term (weeks -- months) variability 
studies. In addition to these previously unpublished data, we included 
publically available photometric monitoring data from the Small and 
Moderate Aperture Research Telescope System 
(SMARTS)\footnote{\tt http://www.astro.yale.edu/smarts/} at the Cerro 
Tololo Interamerican Observatory, in our data collection. Specifically,
we included B, V, and R-band data from the Yale Fermi/SMARTS project, 
covering the entire, extended campaign period. We also included publically 
available polarimetry and spectroscopy
data from the University of Arizona -- Steward 
Observatory {\it Fermi} support 
program\footnote{\tt http://james.as.arizona.edu/~psmith/Fermi/} in
our analysis. 

\subsection{\label{optical}Optical Observations}

\subsubsection{\label{phot}Optical Photometry}

Most of our optical (BVR) observations were carried out  using the 1.3-m 
McGraw-Hill telescope of the MDM Observatory on the south-west ridge of 
Kitt Peak, Arizona. The telescope is equipped with a 1024x1024 pixels CCD 
camera and standard Johnson-Cousins UBVRI filters. Every night during the
period September 9 -- 19, 2009, sequences of science frames on PKS~0528+134 
were taken in the B-V-R filters with exposure times of 360, 180, and 180 
seconds, respectively, with slight variations depending on atmospheric 
conditions. All frames were bias-subtracted and flat-field corrected using 
standard routines in IRAF. In most cases, the signal-to-noise ratio (SNR)
of PKS~0528+134 on individual, reduced frames was too low to allow for a
high-precision magnitude determination. Due to the high redshift of
$z = 2.07$ and the expectation that the time scale (in the cosmological
rest frame of the blazar) of brightness variations in FSRQs is $\gtrsim
1$~hr, we typically added reduced frames taken within intervals no longer 
than three hours in order to increase the SNR. 

In order to construct light curves of this quasar we applied relative 
photometry to the frames using  the three comparison stars in the sky 
chart provided by \cite{raiteri98}. We extracted instrumental magnitudes 
of the three comparison stars and PKS~0528+134 using the {\it phot} routine 
within the DAOPHOT package of IRAF. 
The light curves thus constructed are analyzed in section \S \ref{variability}.

Additional BVRI photometric and R-band polarimetric observations
of PKS~0528+134 were performed at the 1.8~m Perkins telescope of Lowell
Observatory (Flagstaff, AZ) during a 2-week campaign from October
15 to October 28 2009 with the PRISM camera\footnote{http://www.bu.edu/prism/}.
Standard bias subtraction and flat field correction were carried out
for each frame. For these images we also performed differential photometry 
with an aperture of radius 6" using BVR measurements for comparison stars 3, 
2, and 1 from \cite{raiteri98}.  We obtained I-band measurements
for the same comparison stars ($14.475 \pm 0.010$, $13.239 \pm 0.010$, 
and $12.220 \pm 0.010$ mag, respectively) using comparison stars from 
the field of PKS~0735+138 \citep{smith85} observed just after 
PKS~0528+134 on 7 nights. The PRISM camera possesses a polarimeter 
with a rotating half wave plate that we employed for R-band polarimetry. 
Most details of the polarization observations and data reduction can be found 
in \cite{J10}. 
The polarization data were corrected for the statistical bias associated
with the fact that the degree of polarization is restricted to being a positive 
quantity \citep{wk74}.

We also obtained optical data taken using the polarimetric imaging system Polima
attached to the 84~cm telescope at San Pedro M\'artir, Baja California, 
Mexico as part of a long-term polarimetric monitoring program of a sample 
of 35 blazars during 6 photometric nights in October and November 2009.
Due to the optical design of Polima, 4 images with different position angles 
of the polarimeter to measure the Stokes parameters were taken through an 
R-band filter. The data reduction for these frames (correction for bias and 
pixel-to-pixel variations across the CCD) as well as the photometry has 
been carried out on each of the individual images using a dedicated 
pipeline developed by D. Hiriart. The fluxes of PKS~0528+134 and several 
stars in the field of view for each PA have been combined. Flux calibration 
was finally done via the comparison stars 2 and 3 in the field of view from 
\cite{raiteri98}.

\begin{figure}
\plotone{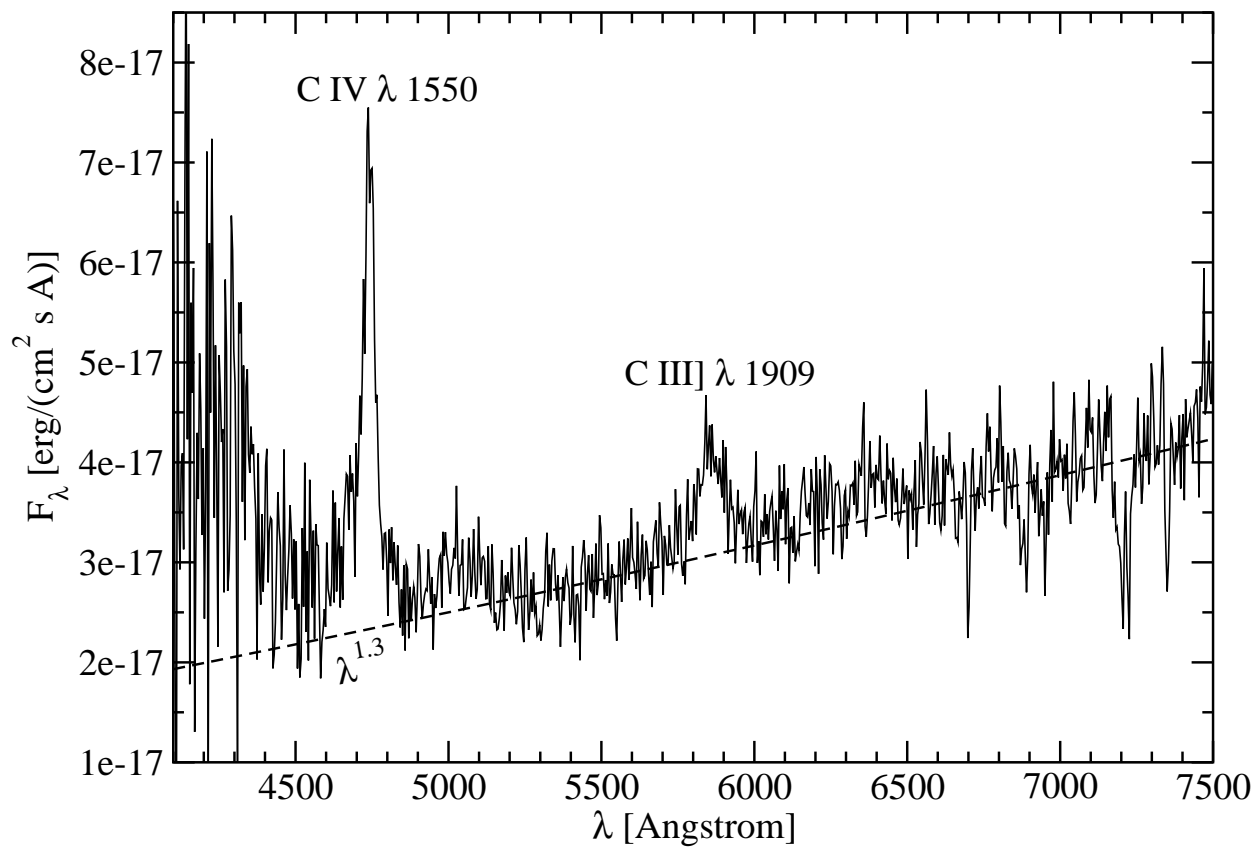}
\caption{Optical spectrum of PKS~0528+134 on December 19, 2009, taken
at the Steward Observatory. The two prominent emission lines of CIII] and
CIV are labeled and correspond to luminosities of $L_{CIII]} \approx 
4.2 \times 10^{44}$~erg~s$^{-1}$ and $L_{CIV} \approx 3.2 \times
10^{45}$~erg~s$^{-1}$. The continuum red-ward of $\sim 5000$~\AA\ can
be fit with a power-law $F_{\lambda} \propto \lambda^{1.3}$.}
\label{Steward_spectrum}
\end{figure}

\subsubsection{\label{spectroscopy}Optical Spectroscopy}

In addition to the photometric observations described in the previous 
sub-section, PKS~0528+134 is also regularly monitored with spectroscopic
observations through the University of Arizona -- Steward Observatory 
{\it Fermi} Support program. As a representative example, we show in
Figure \ref{Steward_spectrum}, the optical spectrum of October 19, 2009,
which is within the extended period of our multiwavelength campaign.
The source was clearly still in the quiescent state targeted in this 
work. 

The spectrum exhibits two distinct emission lines: CIII] $\lambda$1909,
red-shifted to $\lambda = 5860$~\AA, and CIV $\lambda$1550, red-shifted
to $\lambda = 4740$~\AA. The total measured flux in the lines corresponds
to $F_{CIII]} = 1.8 \times 10^{-15}$~erg~cm$^{-2}$~s$^{-1}$ and
$F_{CIV} = 4.3 \times 10^{-15}$~erg~cm$^{-2}$~s$^{-1}$. In order to
evaluate the de-absorbed fluxes, we evaluate the extinction coefficients
$A_{\lambda}$ with the extinction law of \cite{cardelli89} and $A_V =
2.78$, yielding $A_{4740} = 3.35$ and $A_{5860} = 2.59$. This
yields intrinsic luminosities in the two lines of $L_{CIII]} \approx 
5.8 \times 10^{44}$~erg~s$^{-1}$ and $L_{CIV} \approx 2.5 \times
10^{45}$~erg~s$^{-1}$. 

The continuum redward of $\sim 5000$~\AA\ can be well fit with a 
power-law $F_{\lambda} \propto \lambda^{1.3}$, which corresponds
to a steep spectrum in frequency space, $F_{\nu} \propto \nu^{-3.3}$.

\subsubsubsection{\label{polarimetry}Optical Polarimetry}

R-band photo-polarimetric observations of PKS~0528+134 were also acquired with
the 2.2~m telescope of the Calar Alto Observatory in Amer\'\i a, Spain, as part
of the MAPCAT\footnote{Monitoring AGN with Polarimetry at the Calar Alto Telescopes:
http://www.iaa.es/$\sim$iagudo/research/MAPCAT} program. The data were reduced and
calibrated as described in \cite{agudo10} and \cite{J10}. 

St.~Petersburg observations were performed at the 70-cm telescope of 
Crimean Observatory using ST-7~XME photometer-polarimeter. The standard 
procedure, including bias and dark subtraction, flat-field correction 
and calibration relative to comparison stars 1, 2 and 3 from \cite{raiteri98} 
was applied.

\begin{figure}
\plotone{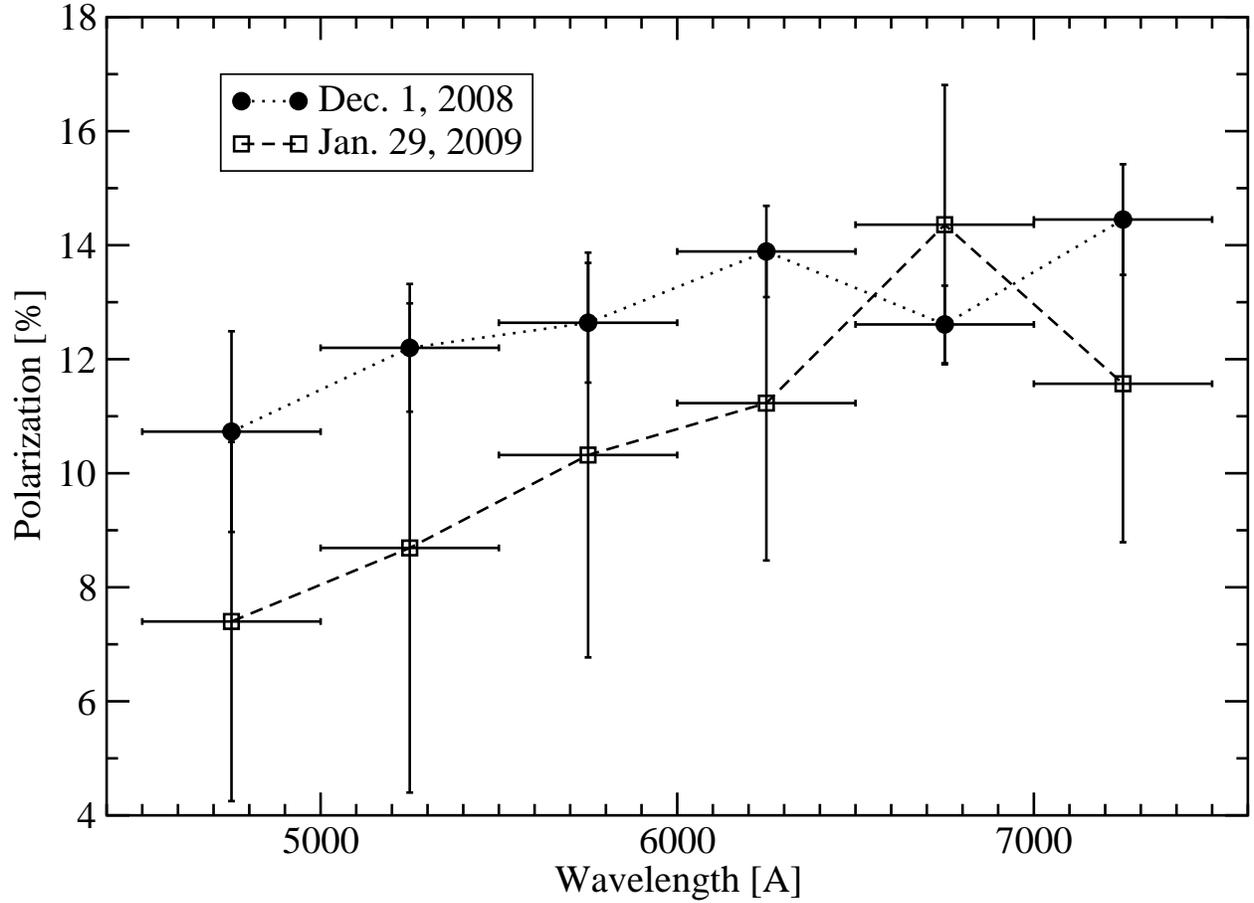}
\caption{Wavelength dependence of the degree of polarization of the
optical emission of PKS~0528+134, from two observations in 2008 December 
and 2009 January. The plot is highly suggestive of a systematic trend of
increasing polarization with increasing wavelength. }
\label{spectropol}
\end{figure}

Additional polarimetry and spectropolarimetry data were included from the 
University of Arizona -- Steward Observatory {\it Fermi} Support program.
Those data included two spectropolarimetric observations of PKS~0528+134 
in 2008 December and 2009 January, which showed a rather high ($P \gtrsim 10$~\%)
degree of polarization, allowing for a meaningful spectrapolarimetric analysis. 
At those times, the object was in a similar quiescent state as during our 
campaign. The contributions of the CIII] and CIV lines (see previous sub-section)
were assumed to be unpolarized, and their fluxes subtracted from the
respective energy bins. Figure \ref{spectropol} shows the degree of
polarization as a function of wavelength. The plot is highly suggestive
of a systematic increase of the polarization towards longer frequencies.
This may be interpreted as an increasing contribution of synchrotron emission
towards the red end of the optical spectrum.

\subsection{\label{xray}X-Ray Observations}

\subsubsection{\label{XMM}XMM-Newton}

PKS 0528+134 was observed by the {\it XMM-Newton Observatory} \citep{Jansen2001}
between September 8th and 14th 2009 on four consecutive revolutions. Table
\ref{pks_epic_data} summarizes these observations. Here, we focus on the data
taken with the EPIC detector, covering the energy band between 0.2 and 10~keV.
EPIC observations were taken in full frame mode and thin optical filter 
(except the pn exposure in 0600121501 that was taken on small window mode). 
{\it XMM-Newton} also has an Optical/UV Monitor Telescope (OM) \citep{Mason2001}, 
a small 30~cm telescope co-aligned with the main XMM-Newton X-ray telescopes. 
However, due to the optical faintness of PKS~0528+134 ($\gtrsim 19^{\rm th}$~mag), 
OM observations did not return useful data. 

\begin{table*}[h]
\centering
\tiny{
\begin{tabular}{cccccccc}
\hline
\hline
Obs. ID  &   Date                & Exp. (pn) & CR (pn)     & Exp. (MOS1) & CR (MOS1)    & Exp. (MOS2)  & CR (MOS2) \\
         &   yy/mm/dd            &  (ksec)   & (cts/sec)   &   (ksec)    & (cts/sec)    &    (ksec)    & (cts/sec) \\
\hline
0600121401&  2009-09-08$@$05:55UT & 9.94(14.28)&0.2675$\pm$0.0052&15.2(22.1)&0.0897$\pm$0.0024&15.17(21.49)&0.0943$\pm$0.0025\\ 
0600121501&  2009-09-10$@$06:05UT & 19.89(20.13)&0.2194$\pm$0.0033&23.63(28.46)&0.0848$\pm$0.0019&24.38(28.48)&0.0897$\pm$0.0019\\ 
0600121601&  2009-09-12$@$03:01UT & 21.11(21.96)&0.2616$\pm$0.0035&26.5(26.5)&0.0808$\pm$0.0017&26.6(26.6)&0.0908$\pm$0.0018\\ 
0600121701&  2009-09-14$@$02:52UT & $---$&$---$&26.68(35.61)&0.0754$\pm$0.0017&29.72(36.12)&0.0877$\pm$0.0017\\ 
\hline
\end{tabular}
}
\caption{Summary of PKS~0528+134 {\it XMM-Newton} EPIC observations. {\it Exp.} =
Livetime after (before) correction due to periods of higher background
activity. {\it CR} = Count rates, given for the 0.2-10~keV energy range. 
The quadrant of the pn chip containing the source of the pn exposure in 
observation 0600121701 was lost and hence there is no source information.}
\label{pks_epic_data}
\end{table*}

\subsubsubsection{\label{XMMreduction}Data Reduction} 

The data have been analysed using SASv9.0 \citep{Gabriel2004} and corresponding 
calibration files. Event and source lists were obtained for the EPIC detector, 
following the standard SAS data reduction procedures. Several filtering criteria 
have been applied. The event list has been filtered for time periods of high 
background activity following the standard procedure of removing those time 
periods with background event rates at energies $E > 10$~keV higher than 
1.0~cts/sec and 0.8~cts/sec for EPIC-pn and EPIC-MOS, respectively. The 
time losses due to the removal of these periods are up to 30~\% depending on 
the observation and the instrument (three out of the four observations were
performed at the end of their respective revolutions with the consequent
increase in radiation levels towards the end of the observations). Table
\ref{pks_epic_data} shows the livetimes for each observation and instrument
before and after this correction has been performed. The data was also
filtered to include only single and double (PATTERN$\leq$4) pattern events for
EPIC-pn and single to quadruple (PATTERN$\leq$12) for EPIC-MOS as well as
those with quality FLAG=0. The filtered event lists were used to generate
light curves and spectral products. The source region was considered to 
be a circular region centered around the source. Annular background 
regions were chosen to be centered around the source with radii 60'' 
$\leq$ R $\leq$ 80''. The source extraction regions were then optimized
based on S/N for the given background region, which yielded typical values of
$\sim$ 40'' in radius for the source region. 

\begin{table}
\begin{center}
\small{
\begin{tabular}{l|cccccc}
\hline
& \multicolumn{5}{c}{EPIC Power Law}\\
& \multicolumn{5}{c}{PHABS+PHABS}\\
\hline
Obs ID
& N$_{\rm H, mol}$
& $\Gamma_{\rm ph}$
& F$_{0.2 - 2 \rm keV}$
& F$_{2 - 10 \rm keV}$
& $\chi^2/d.o.f.$
\\

& 10$^{22}$ cm$^{-2}$
&
& 10$^{-12}$ ergs cm$^{-2}$ s$^{-1}$
& 10$^{-12}$ ergs cm$^{-2}$ s$^{-1}$
&
\\

\hline

0600121401
& 0.15$^{+0.04}_{-0.04}$
& 1.57$^{+0.07}_{-0.07}$
& 0.28$^{+0.03}_{-0.04}$  (0.88)
& 1.35$^{+0.11}_{-0.11}$  (1.39)
& 0.93 (146.2/157)
\\

0600121501
& 0.13$^{+0.02}_{-0.03}$
& 1.60$^{+0.05}_{-0.07}$
& 0.24$^{+0.02}_{-0.03}$ (0.76)
& 1.10$^{+0.07}_{-0.09}$ (1.14)
& 1.02 (243.3/238)
\\

0600121601
& 0.17$^{+0.02}_{-0.02}$
& 1.61$^{+0.05}_{-0.02}$
& 0.24$^{+0.02}_{-0.02}$ (0.81)
& 1.14$^{+0.07}_{-0.06}$ (1.18)
& 1.05 (275.2/261)
\\

0600121701
& 0.12$^{+0.64}_{-0.56}$
& 1.58$^{+0.07}_{-0.10}$
& 0.26$^{+0.03}_{-0.02}$ (0.69)
& 1.18$^{+0.11}_{-0.11}$ (1.22)
& 0.99 (135.6/137)
\\

\hline
\end{tabular}
}
\end{center}
\caption{Fit performed in the 0.2 -- 10~keV energy range. All 3 EPIC 
instruments have been used simultaneously in the fit. The 
Galactic absorption traced by 21-cm emission from neutral hydrogen,
N$_{H, 21cm}$
has been fixed to a value of $0.24 \cdot 10^{22}$~cm$^{-2}$ 
according to the LAB Survey of Galactic HI \cite{Kalberla2005}. Fluxes 
given are not absorption corrected (in parenthesis the de-absorbed flux 
is given). Errors indicate the 90~\% CL. {\it NOTE}: For observation id 
0600121701 only MOS1 and MOS2 are available. In this case, the fit has 
been performed in the 0.3 -- 10~keV energy range and the fluxes given 
are between 0.3 -- 2.0 and 2.0 -- 10 keV.}\label{EPIC_model}
\end{table}

\subsubsubsection{\label{EPICspectra}EPIC Spectral Analysis}

Time-averaged spectra have been obtained for each individual
observation. The spectra were re-binned in order not to oversample the
intrinsic energy resolution of the EPIC cameras by a factor larger than 3,
while making sure that each spectral channel contains at least 25
background-subtracted counts. Both conditions allow the use of the $\chi^2$
quality-of-fit estimator to find the best fit model. Fits were performed in
the 0.2 -- 10~keV energy range. The spectra from all three EPIC cameras have 
been used simultaneously during the fitting procedure. The systematic difference
between the EPIC cameras is below $\sim 5 $~\% in normalization. For the
spectral analysis and fitting procedure, XSPEC v12.4 \citep{Arnaud1996} was
used.

The spectral model used to fit the data is composed of a power law
convolved with a combination of Galactic column density of absorbing
material traced by 21-cm emission from atomic hydrogen (N$_{\rm H, 21cm}$),
plus an additional absorbing column density (N$_{\rm H, mol}$), primarily 
due to heavy elements in the intervening molecular cloud B30, which are not 
properly traced by the hydrogen column density. The Galactic hydrogen 
column density is kept fixed during the fitting procedure. The spectral 
fitting model takes the following form for the differential photon 
flux $\Phi(E)$:

\begin{center}
\begin{equation}
\Phi(E) = e^{- N_{\rm H, 21cm}\sigma(E)} \cdot e^{-N_{\rm H, mol}\sigma(E)} \cdot 
N \cdot E^{-\Gamma_{\rm ph}} 
\end{equation} 
\end{center}

where $\sigma(E)$ is the photoelectric absorption cross-section, with 
abundances after \cite{Anders1989}, and $\Gamma_{\rm ph}$ the power-law 
index with normalization {\it N}. Errors in the relevant parameters are 
given at the 90~\% confidence level (CL) for any given parameter.

\begin{figure}
  \centering
  \includegraphics[width=.75\textwidth,angle=270]{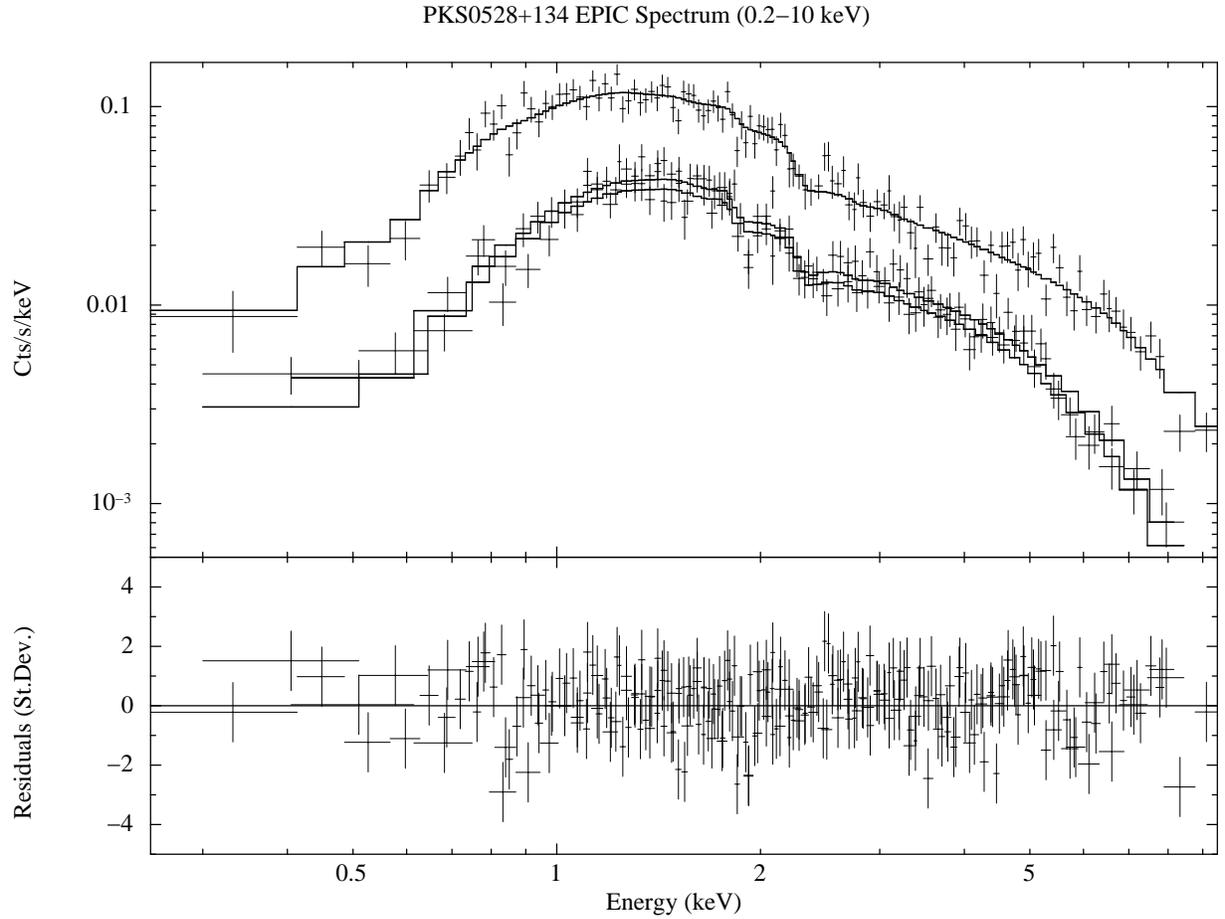}\\
  \caption{Example of EPIC (pn, MOS1 and MOS2) spectrum for the observation id
  0600121601 where EPIC data is available for all three cameras. The fit has
  been done in the 0.2-10~keV energy range with channel grouping $> 3$
  channels and $> 25$ background subtracted counts.}
  \label{epic_spectrum}
\end{figure}

Table \ref{EPIC_model} shows the best fit values for the model considered
for all four observations. Figure \ref{epic_spectrum} shows an example of the
spectra as determined for one of the observations (composed of EPIC-pn and the
two EPIC-MOS spectra) with the best fit model and $\chi^2$ deviation. Similar
spectra are derived for the other observations.

\subsubsection{\label{RXTE}RXTE}

In addition to our four ToO {\it XMM-Newton} pointings, we monitored 
the 2.4 -- 10~keV X-ray flux of PKS~0528+134 with the {\it Rossi
X-ray Timing Explorer} (RXTE) Proportional Counter Array (PCA), 
with exposure times of 900 -- 3400~s for individual observations.  
The X-ray flux measurement entailed the subtraction of an X-ray 
background model (faint source model, version 20051128) from the 
raw spectrum using the standard X-ray data analysis software 
packages FTOOLS and XANADU. We used the program XSPEC to fit the
residual photon spectrum with a power-law model plus photoelectric
absorption along the line of sight.

\subsubsection{\label{Suzaku}Suzaku}

PKS~0528+134 was also observed with the Suzaku satellite as a part of 
multi-band observations conducted in September 2008, about one year prior 
to the campaign covered in this paper.  However, both the X-ray flux and 
$\gamma$-ray flux levels were comparable to those measured during the
September 2009 campaign, indicating a quiescent state. We therefore include
the {\it Suzaku} data here for comparison of the spectral and short-term
variability properties with those measured by {\it XMM-Newton} one year later, 
in a similar quiescent state.
The {\it Suzaku} satellite features instruments 
sensitive in the soft X-ray band.

Suzaku observations of PKS~0528+134 started on 2008 September 27, 02:38 UT, 
and lasted until 2008 October 2, 16:12 UT.  Since the source was not 
detected in the Hard X-Ray Detector (HXD) data, we considered only the 
X-Ray Imaging Spectrometer (XIS) instruments.  
The observation conditions were nominal, although the XIS1 data 
suffered from unusually high and variable background, resulting in a 
total apparent background count rate ranging from 1 to 3 counts~s$^{-1}$ 
over the entire chip.  Nonetheless, since the background-subtracted 
spectrum determined from the XIS1 data was entirely consistent with 
that from XIS0 and XIS3, we included the properly background-subtracted 
XIS1 data in the spectral fitting (see below). 

\begin{figure}
  \centering
  \includegraphics[width=.75\textwidth,angle=270]{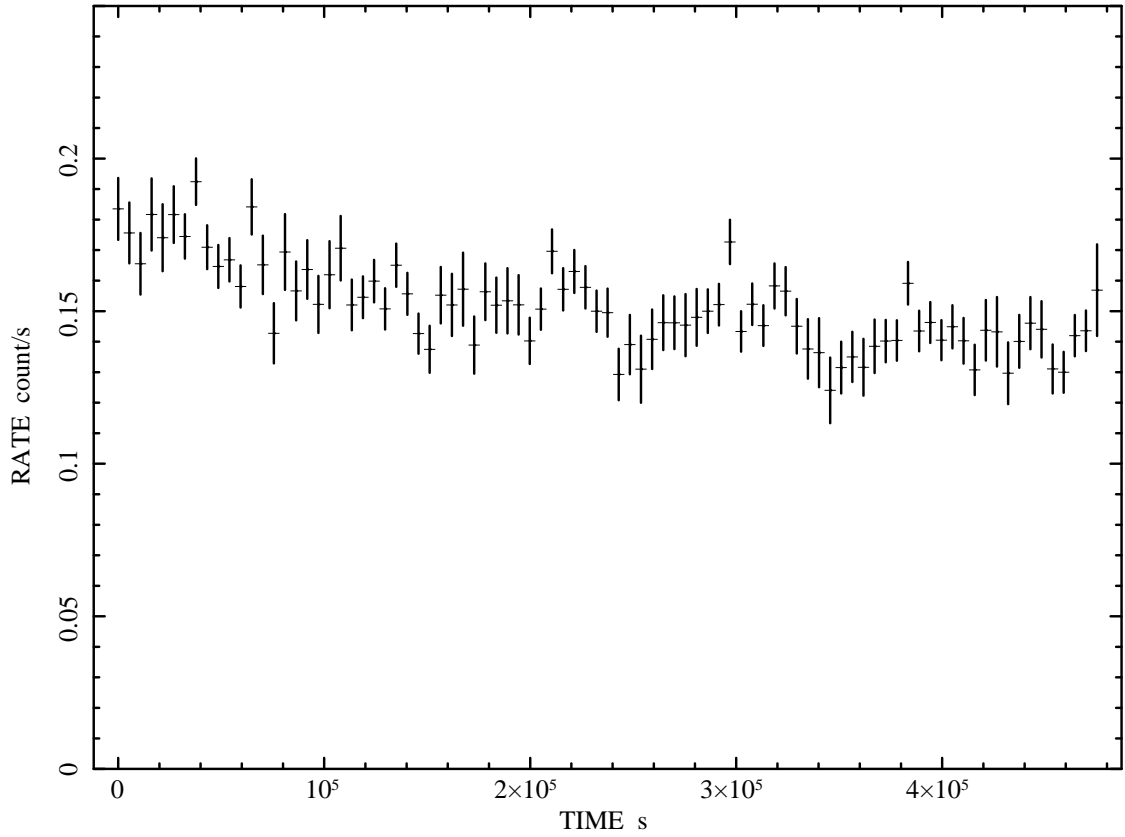}\\
  \caption{{\it Suzaku} X-ray light curve of PKS~0528+134 from the XIS3 data 
  of 2008 September 27 -- October 2. 
  $t = 0$ corresponds to MJD~54736.1771.
  }
  \label{Suzaku_lc}
\end{figure}

The total exposure time yielding good data accumulated in the pointing was 
203~ks. We used the standard {\tt ftools} data reduction package, provided
by the Suzaku Science Operations Center, with the calibration files 
included in the CALDB ver. 4.3.1.  The net count rates were, respectively, 
0.10, 0.13, and 0.12 count~s$^{-1}$ for XIS0, XIS1, and XIS3.  
For the analysis of spectra and light curves, we extracted the 
counts from a region corresponding to a circle with 260 arc sec 
radius. We used a region of a comparable size from the same chip 
to extract the background counts.  We plot the resulting total light 
curve (not background-subtracted) from XIS3, binned in 5400 sec bins, 
in Figure \ref{Suzaku_lc}, and note that the background was steady at the 
level of $\sim 0.015$ count~s$^{-1}$.  The data indicate {\sl no significant 
rapid (a day or less) variability} during the {\it Suzaku} observation, 
but show a {\sl long-term trend}, with a $\sim 20$~\% decrease in flux 
over the $\sim 5.3$~day long duration of the observation. A similar 
secular flux decrease is seen in the two other XIS detectors.

\begin{figure}
  \centering
  \includegraphics[width=.75\textwidth,angle=270]{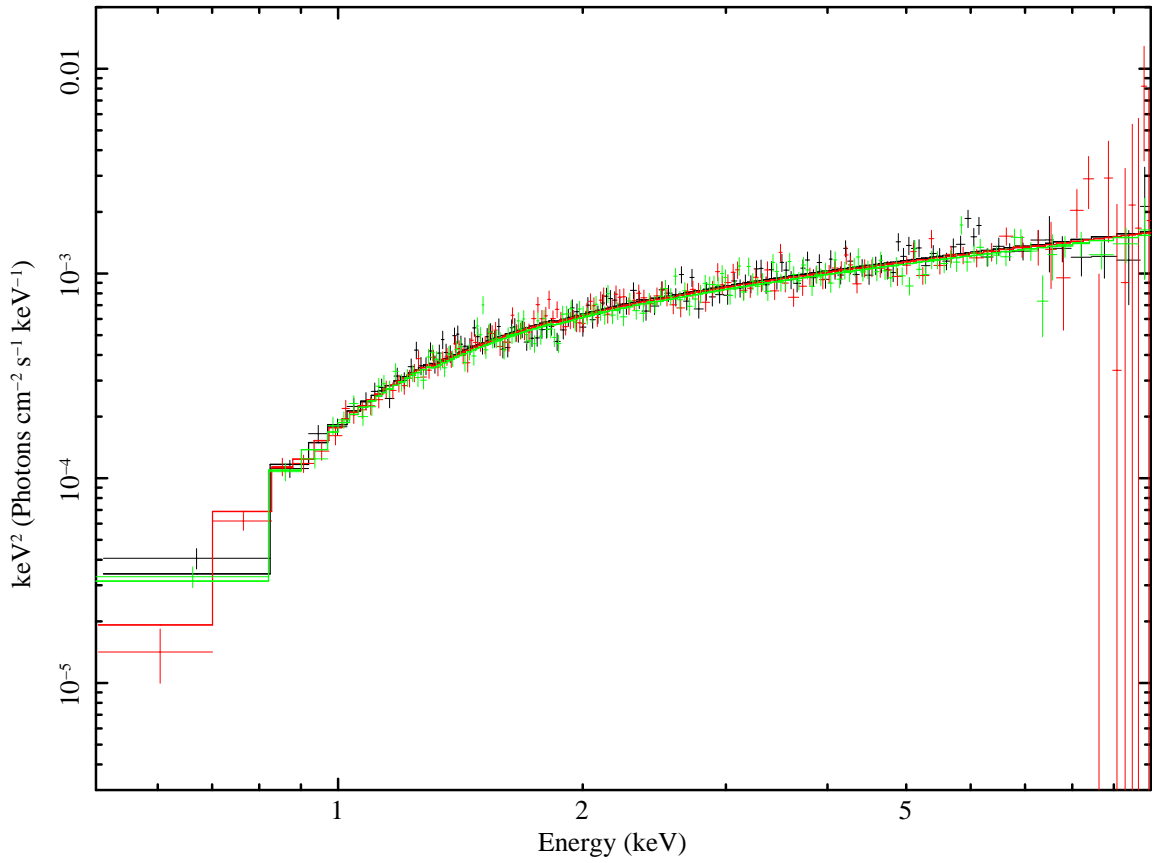}\\
  \caption{0.5 -- 10~keV X-ray spectrum of PKS~0528+134 from the {\it Suzaku}
  observations of 2008 September 27 -- October 2.}
  \label{Suzaku_sp}
\end{figure}

For the spectral analysis, we we used the {\tt XSPEC} spectral analysis software 
\citep{Arnaud1996}. For the spectral fitting, we used the standard redistribution 
files and mirror effective areas generated with {\it Suzaku}-specific tools.  
We included the counts corresponding to the energy range of 0.5 -- 10.0~keV in 
our spectral fits.  We used all three XIS detectors simultaneously, but allowed 
for a small (a few \%) variation of the relative normalizations.  
The source spectrum was modeled as an absorbed power law, with the 
cross-sections and elemental abundances as given in \cite{mmc83}.
Other absorption models give similar results. The best-fit absorbing 
column was $4.6 \pm 0.2 \times 10^{21}$ cm$^{-2}$, and the photon index 
was $1.54 \pm 0.03$, with the acceptable best-fit $\chi^{2}$ of 3407 for 
3512 d.o.f.  The {\it Suzaku} spectrum is shown in Figure \ref{Suzaku_sp}.
The observed model 2 -- 10 keV flux is $2.8 \times 10^{-12}$~erg~cm$^{-2}$~s$^{-1}$, 
with the statistical error of 5~\%, which is probably smaller than the systematic 
error resulting from the calibration uncertainty of the {\it Suzaku} instruments.  

The fitted values of the total absorbing column are only marginally 
consistent between the {\it Suzaku} and {\it XMM-Newton} data sets. This
might be in part related to the differences in calibration of the two 
instruments. An additional source for the discrepancy might be the fact
that (1) we are using Galactic elemental abundances, which might or might 
not be appropriate for the line-of-sight molecular cloud, and (2) the
bandpass of {\it XMM-Newton} extends down to 0.3~keV, while for Suzaku
it is 0.5~keV. If the elemental abundances in the intervening absorber
of low-z elements are not the same in the molecular cloud as the Galactic
abundances assumed in the spectral fit, the precise value of the best-fit
$N_H$ might be dependent on the bandpass of the instrument and even the 
details of the effective area at the energies where the absorption edges 
due to those elements play an important role.

While the X-ray spectral index measured by {\it Suzaku} in 2008 September/October
is consistent with the value measured one year later by {\it XMM-Newton} (see
\S \ref{XMM}, table \ref{EPIC_model}), the {\it Suzaku} 2 -- 10~keV flux is 
about a factor 2.4 higher than the {\it XMM-Newton} value. This is in agreement
with the $\sim$~factor 2 -- 3 variability measured by {\it RXTE} on $\sim$~weekly
time scales (see \S \ref{fluxvar}).

\subsection{\label{gammaray}$\gamma$-Ray Observations}

Gamma-ray observations of PKS~0528+134 were performed by the {\it Fermi}-LAT. 
This is a pair-conversion $\gamma$-ray telescope sensitive to photon energies 
greater than 20 MeV. In its nominal scanning mode, it surveys the whole sky every 
3 hours with a field of view of about 2.4 sr \citep{atwood09}. The data presented 
in this paper (restricted to the 100~MeV -- 200~GeV range) were collected from 
MJD~54983 (2009 June 1st) to MJD~55193 (2009 December 28). For this analysis, 
the {\it Diffuse} event class was used. This is the optimized class for point 
source analysis with minimal residual contamination from charged-particle 
backgrounds. To minimize systematics, only photons with energies greater 
than 100~MeV were considered in this analysis. In order to limit the 
contamination from atmospheric $\gamma$-rays produced by interactions 
of cosmic rays with the upper atmosphere of the Earth, only events with 
zenith angle $<$ 105$^{\circ}$ were selected. In addition, time intervals 
during which the rocking angle was larger than 52$^{\circ}$ have been excluded 
from the analysis, because the bright limb of the Earth enters the field of 
view. The analysis was performed with the standard analysis tool {\it gtlike}, 
part of the Fermi-LAT Science Tools software package (version 
v9r15p5)\footnote{For a documentation of the Science Tools see
http://fermi.gsfc.nasa.gov/ssc/data/analysis/documentation/}. The
P6\_V3\_DIFFUSE set of instrument response functions was applied.

Photons were selected in a circular Region Of Interest (ROI) 
7$^{\circ}$ in radius, 
centered at the position of PKS~0528+134. The isotropic background, including 
the sum of residual instrumental background and extragalactic diffuse
$\gamma$-ray background, was modeled by fitting this component at high 
galactic latitude (file provided with Science Tools). The Galactic and 
Isotropic diffuse emission models version ``gll\_iem\_v02.fit'' and 
``isotropic\_iem\_v02.txt'' were used.\footnote{For details on the background 
model see http://fermi.gsfc.nasa.gov/ssc/data/access/lat/BackgroundModels.html} 
All point sources in the first Fermi-LAT catalog \citep{abdo10c}
lying within the ROI and a surrounding 
10$^{\circ}$-wide annulus were considered in the fit and modeled with
single power-law distributions of the form $F(E) = N_0 (E/E_0)^{-\Gamma}$. 
Their flux was kept free whereas their spectral index value was frozen to 
the value listed in the 1FGL catalog, except for PKS~0528+134 whose 
index was kept free. Due to limited statistics, the gamma-ray spectrum of 
PKS~0528+134 was modeled with a power-law distribution in the present
analysis, although the spectrum measured over 11 months 
covered by the 1FGL catalog exhibits distinct  
curvature, characterized by a curvature index of 8.14, corresponding to a 
4~\% probability that the spectrum is adequately represented by a power law 
\citep{abdo10}. The highest energy photon attributed to PKS~0528+134 has 
an energy of 8.6~GeV. The estimated systematic uncertainty on the flux 
is 10~\% at 100~MeV, 5~\% at 500~MeV and 20~\% at 10~GeV.

\subsection{\label{radio}Radio Observations}

We obtained radio observations from the GLAST-AGILE Support Program 
(GASP) of the Whole Earth Blazar Telescope (WEBT). In table \ref{radiodata}, 
we list the frequencies in which we collected data and the observatories 
participating in the campaign.

\begin{deluxetable}{cccccc}
\tabletypesize{\scriptsize}
\tablecaption{Frequencies and observatories used to collect radio data}
\tablewidth{0pt}
\tablehead{
\colhead{Frequency} & \colhead{Observatory}
}
\startdata
05   GHz &         UMRAO, MEDICINA\\
08   GHz &         UMRAO, MEDICINA\\
14   GHz &	UMRAO\\
22   GHz &	MEDICINA\\
37   GHz &	MetsŠhovi Radio Observatory (KURP-GIX)\\
43   GHz &	Noto\\
230 GHz &	MAUNA KEA (SMA)\\
345 GHz &	MAUNA KEA (SMA)\\
\noalign{\smallskip\hrule\smallskip}
\enddata
\label{radiodata}
\end{deluxetable}

Centimeter-band observations were obtained with the University of  Michigan 26-meter
prime focus paraboloid equipped with radiometers operating at central frequencies of
4.8, 8.0, and 14.5~GHz. Observations at all three frequencies utilized rotating
polarimeter systems permitting both total flux density and linear polarization to 
be measured. A typical measurement consisted of 8 to 16 individual pointings
over a 20 -- 40 minute time period. Frequent drift scans were made across stronger
sources to verify the telescope pointing correction curves. Observations of
program sources were intermixed with observations of a grid of calibrator sources
to correct for temporal changes in the antenna aperture efficiency. The flux scale
was based on observations of Cassiopeia A \citep[see, e,g.,][]{baars77}. Details of 
the calibration and analysis techniques are described in \cite{aller85}.

The 37~GHz observations were made with the 13.7~m diameter Mets\"ahovi radio
telescope. The detection limit of the telescope at 37~GHz is of the order
of 0.2~Jy under optimal conditions. Data points with a SNR $< 4$ are handled 
as non-detections. The flux density scale is set by observations of DR~21. 
Sources NGC~7027, 3C~274 and 3C~84 are used as secondary calibrators. A 
detailed description of the data reduction and analysis is given in 
\cite{terasanta98}. The error estimate in the flux density includes the 
contribution from the measurement rms and the uncertainty of the absolute 
calibration.

The 43~GHz Noto observations have been performed using the On The Fly (OTF) 
scan technique (scan duration about 20~s) and the telescope gain (K/Jy) was 
determined as a function of the elevation using NGC~7027 as primary calibrator. 
To improve the signal to noise ratio many scans have been acquired and then 
averaged for a total integration time of about 15 minutes. The antenna 
temperature has been estimated by a gaussian fit of the average scan. 
A more detailed description for the 43~GHz data is given by \cite{leto09}.
Radio observations at the Medicina radio observatory were performed at 5,
8, and 22~GHz, and were analyzed as detailed in \cite{bach07}.

Data at 230~GHz and 345~GHz were collected at the Submillimeter Array 
(SMA) on Mauna Kea and reduced using the MIR data reduction software. 
SMA flux density measurements are produced from a mixture of dedicated 
flux calibration/monitoring observations and data from science projects 
that may utilize a quasar as a calibration standard (typically for gain 
calibration of the interferometer). Raw visibility data are calibrated to 
correct for atmospheric absorption and instrumental gain variations, and then 
referenced to observations of standard flux calibration sources, generally 
planets and/or moons.  More details on the SMA flux density monitoring program 
can be found in \cite{gurwell07}.

\section{\label{variability}Variability Analysis}

One of the main goals of this multiwavelength campaign was the search for
flux and spectral variability of PKS~0528+134 in its quiescent state.
We first describe our results on flux variability in \S \ref{fluxvar} and
then turn to the investigation of spectral variability in \S \ref{spectvar}.
In \S \ref{polvar} we discuss variability of the optical polarization.

\subsection{\label{fluxvar}Flux Variability}

\begin{figure}
\plotone{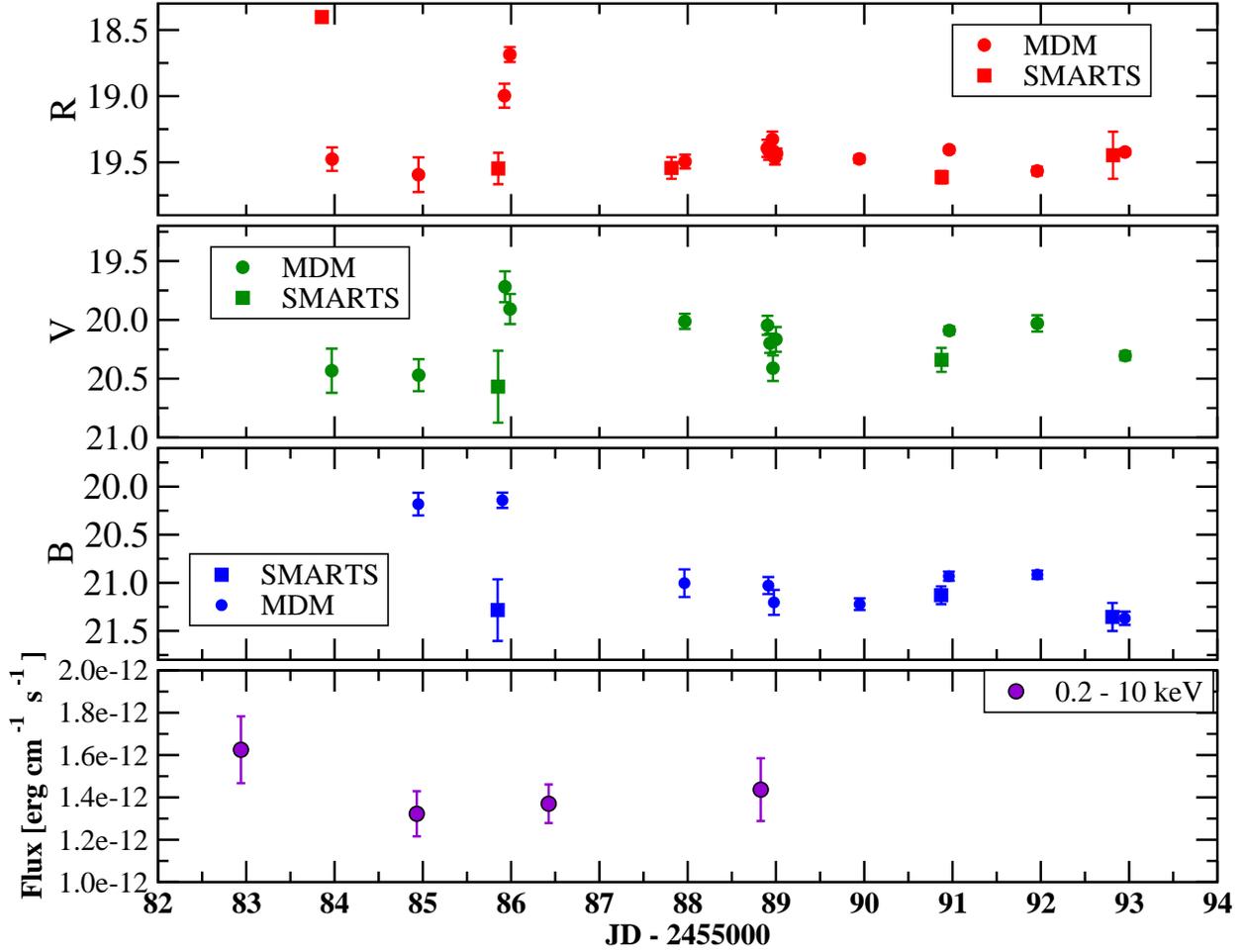}
\caption{MDM 
and SMARTS optical light curves (RVB) of PKS 0528+134 
between September 9 and 19, 2009. The  bottom panel shows  the X-ray light 
curve corresponding to the four {\it XMM-Newton} observations.}
\label{ox_lc}
\end{figure}

The optical light curves of PKS~0528+134 obtained with the 1.3-m McGraw-Hill 
telescope of the MDM Observatory during the core of our multiwavelength campaign  
(September 9 -- September 19, 2009),
along with the publically available SMARTS data from the same period,
are plotted in figure \ref{ox_lc}. An increase 
in the brightness of the source of $\sim$ 0.7 magnitudes in the R and V 
filters around JD 2455086 (September 11) is evident. In order to quantify 
the variability, we computed the reduced $\chi^2$ for a fit to a constant flux. 
Variability is evident in all three bands with $\chi_{\nu}^2 = 15.9$, $5.2$,
and $26.2$, respectively, for the R, V, and B bands. Due to the limited 
observability period of PKS~0528+134 in any given night (typically 
$\lesssim 4$~hr), the light curve during this mini-flare is clearly 
undersampled. Due to the sparse sampling of the optical observations,
we can only constrain the variability time scale to $t_{\rm var}^{\rm opt} 
\lesssim 1$~day.

\begin{figure}
\plotone{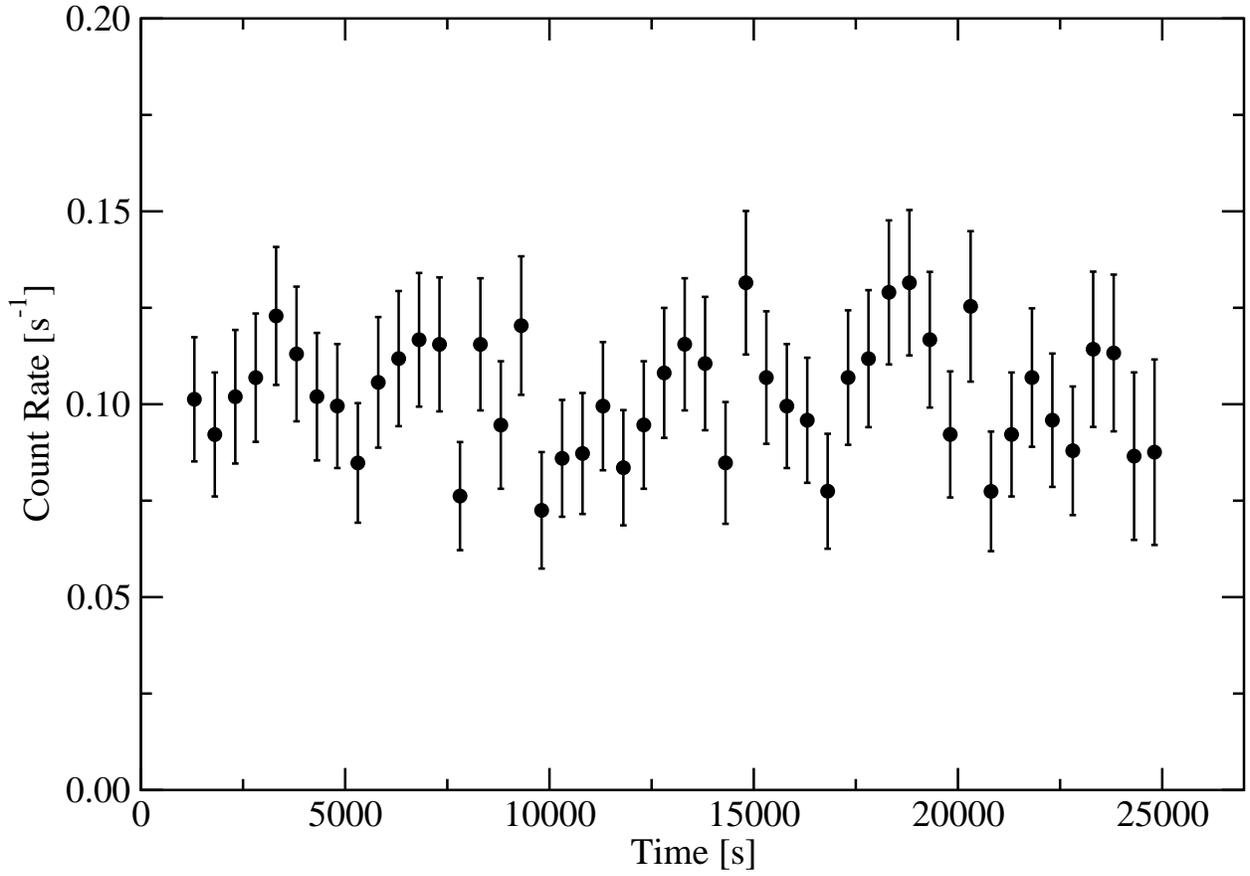}
\caption{{\it XMM-Newton} MOS1 light curve of our ToO observation on 
September 10, 2009. Each point corresponds to a bin of 500 seconds. 
No evidence for intra-day variability is found. }
\label{xray_lc}
\end{figure}

In the same way, we analyzed the X-ray variability 
on time scales of $\sim 1$ -- 2~day between our four {\it XMM-Newton} 
observations as well on time scales of a few hours, within the
individual observations. The {\it XMM-Newton} light curve 
is plotted in the bottom panel of figure \ref{ox_lc}. No flux variability 
is evident. For the 0.2 -- 10~keV X-ray flux, a fit to a constant 
results in a $\chi_{\nu}^2$ of 0.91, thus confirming the absence of 
significant variability. For the same {\it XMM-Newton} observations, we 
also analyze the intraday variability. For most observations the $\chi_{\nu}^2$ 
for a fit to a constant flux is less than 1. In figure \ref{xray_lc}, we show 
the light curve from the {\it XMM-Newton} MOS1 data of September 10, 2009, 
as an example. For this particular observation, $\chi_{\nu}^2 \approx 0.81$ 
is found. 

\begin{figure}
\plotone{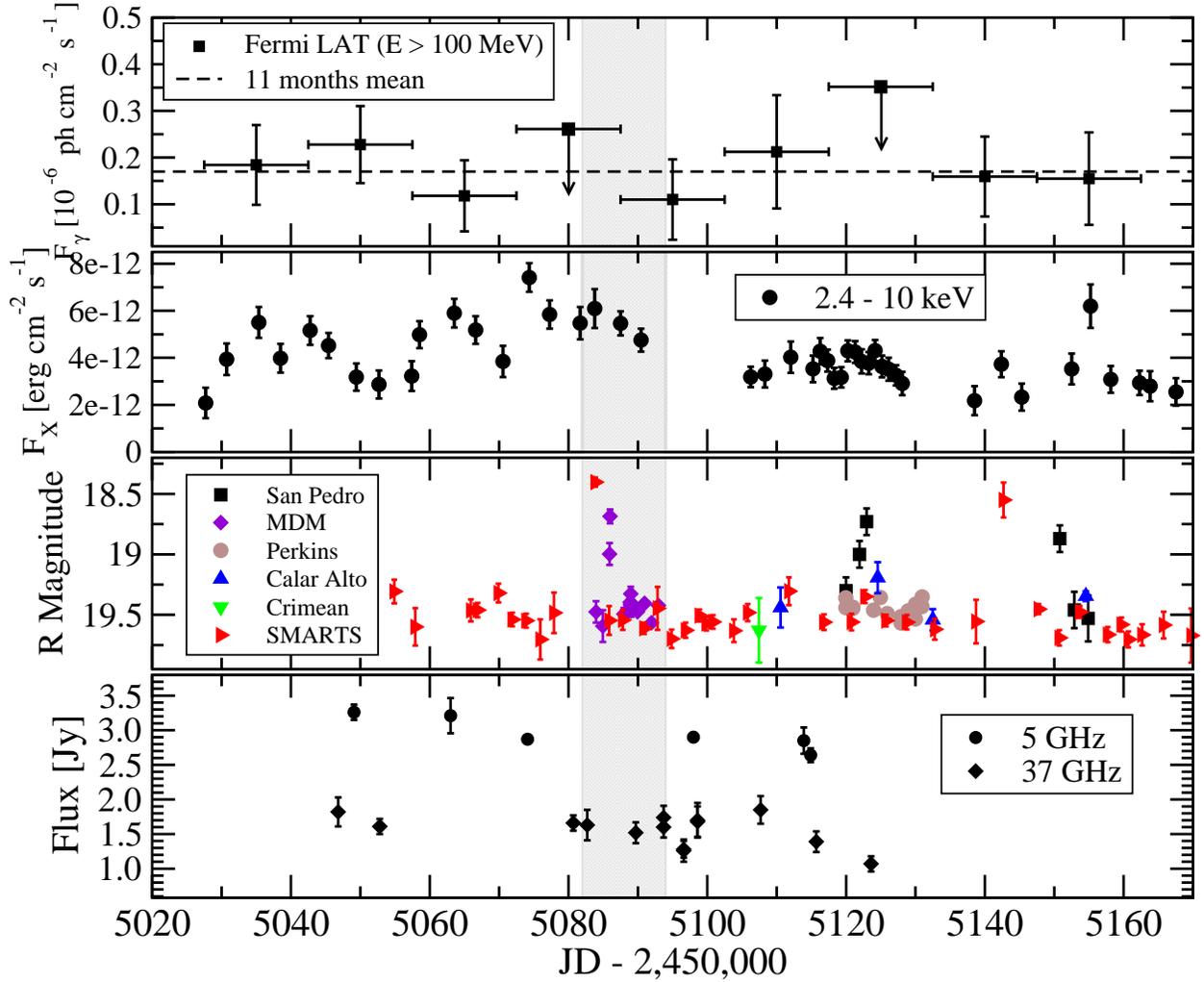}
\caption{Light curves of PKS~0528+134. From top to bottom: a) the Fermi 
$\gamma$-ray flux in 15-day integration bins. b) X-ray (RXTE) light 
curve. c) Optical (R-magnitude) light curve including data from the 
MDM, SMARTS, 
Perkins, San Pedro M\'artir, Calar Alto, and Crimean observatories. 
d) Radio light curves at 5 and 37 GHz. }
\label{multi}
\end{figure}

A multiwavelength variability analysis for a more extended period of time 
was made possible by including RXTE monitoring data. In figure \ref{multi}, 
we show the light curves of PKS~0528+134 in radio (5 and 37~GHz), optical, 
X-rays (RXTE), and $\gamma$-rays (bottom to top). The vertical shaded band 
highlights the interval of the core campaign where the most intense 
optical and X-ray observations (including {\it XMM-Newton}) were 
performed. The RXTE X-ray data from the period July 14 to December 2, 2009, 
represent the most extended coverage. In this interval, PKS~0528+134 shows 
significant variability corresponding to a reduced $\chi_{\nu}^2$ of 3.83
for a fit to a constant flux. The RXTE light curve indicates variability
with flux changes of $\vert \Delta F / F \vert \sim 50$~\%
on a characteristic time scale of $t_{\rm var}^X \sim 1$~week.

We note here that the fluxes resulting from our {\it RXTE} analysis are
systematically higher than those measured by {\it XMM-Newton} during the
same period. We have carefully double- and cross-checked both analyses and 
confirmed this discrepancy. The ROSAT All Sky Survey Catalogue does not list 
any known source in the field of view of {\it RXTE} which may be responsible 
for the flux discrepancy. Furthermore, three {\it RXTE} observations were 
carried out within a few hours of one of the {\it XMM-Newton} observations. 
Given this short time period and the systematic offset between the two 
instruments, rapid variability appears unlikely to cause the flux discrepancy.
The factor of 4 discrepancy seems also too large to be solely due to calibration
uncertainties. A plausible explanation for the systematic offset could lie
in an uncertain background model for the {\it RXTE} analysis in the region
around PKS~0528+134, near the Galactic anti-center. While this would cause
a constant offset of the {\it RXTE} flux levels, it will not affect the
variability. We are therefore confident that the {\it RXTE} flux variability 
analysis presented here is robust. For the analysis of the spectral energy
distributions in \S \ref{SED}, the flux and spectral information from
{\it XMM-Newton} will be used. 

The optical light curve (second panel from the bottom of Figure \ref{multi}), 
including data from the MDM, SMARTS,
Perkins, San Pedro M\'artir, Calar Alto, and 
Crimean observatories, covers the interval from September 9 to November 19, 
2009. This is the band where PKS~0528+134 shows the most significant variations 
in its flux density, with flaring episodes exhibiting brightness changes of 
$\Delta R \lesssim 1^{\rm mag}$ on a time scale of 
$\lesssim 1$~day.

As shown in figure \ref{multi} (top panel), the Fermi $\gamma$-ray flux does 
not present significant variability during the extended campaign period (2009 
July -- December). The best fit found for the flux (E $>$ 100 MeV) is 
$(0.12 \pm 0.02) \times 10^{-6}$~ph~cm$^{-2}$~s$^{-1}$
with a test statistic TS = 68.
The Test Statistic \citep{mattox96} is defined as twice the difference 
in log(likelihood) obtained by including the source of interest and omitting 
it in the source model used in the gtlike analysis.
This flux is slightly lower than the mean 
flux observed over the first eleven months of Fermi data \citep{abdo10}, 
which corresponds to $(0.17 \pm 0.01) \times 10^{-6}$~ph~cm$^{-2}$~s$^{-1}$ 
as represented by the dashed horizontal line in the same panel.

Among the frequencies monitored in the radio regime, the 5 GHz and 37~GHz light 
curves had the most extended coverage in time. In the bottom panel of figure 
\ref{multi}, we show both light curves. Only moderate variability 
with amplitudes of $\vert \Delta F / F \vert \lesssim 20$~\%
is observed at these frequencies. In particular, a decreasing flux tendency is 
found  at the end of the 37~GHz light curve. A fit to a constant flux results in 
$\chi_{\nu}^2 = 3.92$ and 2.61 for the 5 GHz and 37 GHz light curves, respectively. 

\begin{figure}
\plotone{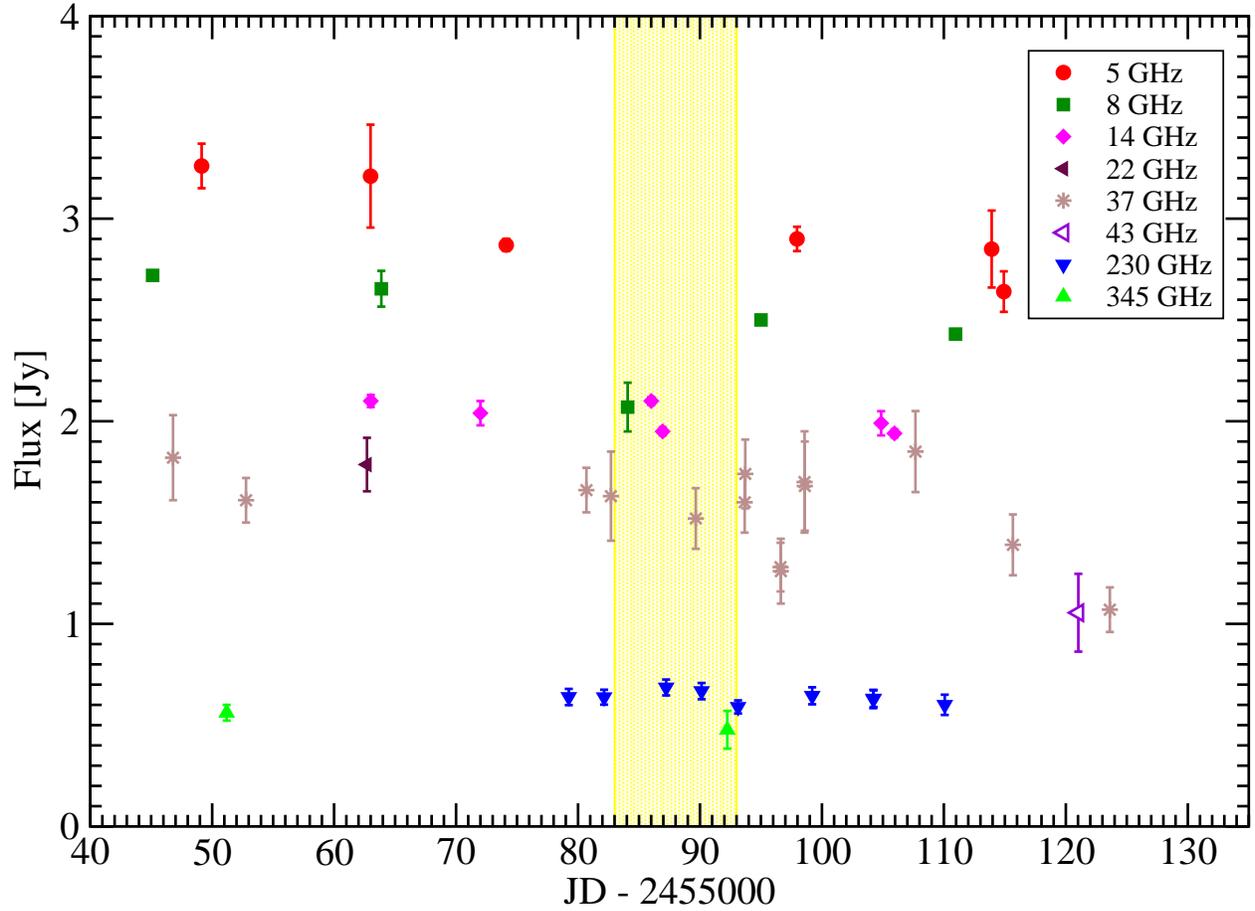}
\caption{Radio light curves of PKS 0528+134. The vertical yellow shaded band 
corresponds to the interval when MDM optical and {\it XMM-Newton} X-ray 
observations of this source were performed.}
\label{radio_lc}
\end{figure}

A more detailed analysis of the radio regime, including the light curves 
in all the monitored radio frequencies, is presented in figure \ref{radio_lc}. 
The radio light curves show moderate variability in general. The 8~GHz 
and 14~GHz bands present significant flux variations with $\chi_{\nu}^2 = 32.96$ 
and 10.17, respectively, for a fit to a constant flux. The results of a 
similar analysis for all radio frequencies are summarized in Table 
\ref{radiovar}. Variability appears to occur on time scales of 
$t_{\rm var}^{\rm radio} \sim 1$ -- 2 weeks, comparable to the 
X-ray variability time scale measured by RXTE (see above).

\begin{deluxetable}{cccccc}
\tabletypesize{\scriptsize}
\tablecaption{Light curve variability analysis in radio frequencies:
$\chi_{\nu}^2$ for fits to a constant flux.}
\tablewidth{0pt}
\tablehead{
\colhead{Frequency} & \colhead{$\chi_{\nu}^2$ }
}
\startdata
5   GHz &         3.92\\
8   GHz &         32.96\\
14  GHz &	  10.17\\
37  GHz &	  2.61\\
230 GHz &	  0.62\\
\noalign{\smallskip\hrule\smallskip}
\enddata
\label{radiovar}
\end{deluxetable}

In summary, the comparative flux variability analysis of PKS~0528+134 at different 
frequencies from $\gamma$-rays through radio indicates no significant variability 
in the $\gamma$-ray regime, moderate variability in the X-rays 
($\vert \Delta F / F \vert \sim 50$~\%) 
and most radio frequencies ($\vert \Delta F / F \vert \lesssim 20$~\%)
on time scales of $\sim 1$ -- 2~weeks, and variability 
of up to $\Delta R \lesssim 1^{\rm mag}$
in the optical bands on time scales of several hours. Our data are
not sampled densely enough for a meaningful analysis of time lags among different 
bands.

\begin{figure}
\centerline{\includegraphics[width=.85\textwidth]{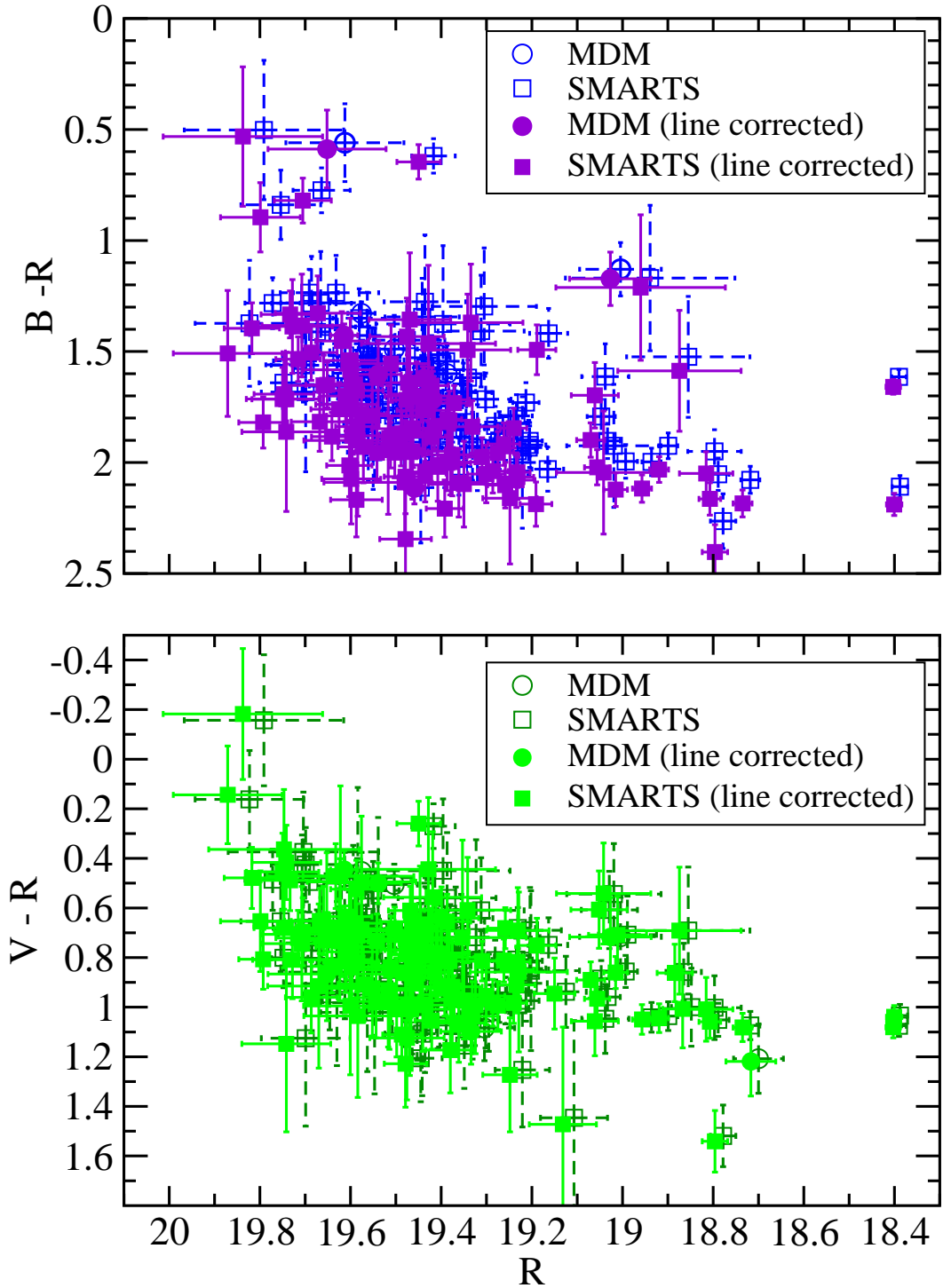}}
\caption{Color-magnitude diagrams for B - R (top panel) and V - R (bottom
panel) vs. R-band magnitude. 
Open symbols with dashed error bars indicate
the original data; filled symbols with solid error bars show the data data
after correction of all magnitudes for the contributions from the CIII] and
CIV emission lines. }
\label{color}
\end{figure}

\subsection{\label{spectvar}Spectral Variability Analysis}

In order to test whether the optical flux variability discussed in the previous 
section is associated with spectral changes, we evaluated color indices B - R 
and V - R as a measure of spectral hardness. These were calculated for any 
pair of B (V) and R magnitudes measured quasi-simultaneously, i.e., during 
the same night, by MDM and SMARTS. 
If conditions were good enough to extract more than one high-quality (error 
in magnitude $< 0.1$) V or B and R band data point per night, magnitude 
measurements taken within $< 1$~hr of each other were used to calculate 
color indices. The resulting flux -- spectral hardness correlations, indicated 
by the color-magnitude diagrams (color vs. R magnitude), are shown 
by the open symbols with dashed error bars in Figure 
\ref{color}. The data clearly indicate color variability. Specifically, 
a weak softer-when-brighter trend seems to be present in both diagrams. 

A correlation analysis of the color-magnitude data sets yields a Pearson's 
correlation coefficient of $r = -0.50$, for the V - R color vs. R magnitude
correlation, and $r = -0.45$ for the B - R color. In both cases the obtained 
correlation coefficients indicate a weak negative correlation between color 
and magnitude, which confirm the redder-when-brighter trend. 
In order to assess the significance of these color correlations, we
performed Monte-Carlo simulations of 10 million data sets with the same
number of data points and the same spread in values as our data. The
R magnitudes and color indices are assumed to be randomly distributed 
and not correlated to each other. For each simulated data set, the 
Pearson's correlation coefficient was evaluated in the same way as 
done for our observational data. This resulted in a chance 
probabilities of $P (< r) = 5 \times 10^{-6}$ and $P ( <r) = 2 
\times 10^{-7}$, respectively, of obtaining a correlation coefficient 
more negative than the ones resulting from our observational data.
This seems to provide strong evidence for the presence of a redder-when-brighter
trend in PKS~0528+134. 

Such a redder-when-brighter trend has been observed in other FSRQs, e.g., 
3C~454.3 \citep{raiteri08}, 
where it has been partially attributed to a contribution of
emission lines from the BLR to the B band. In order to test whether
this may also be the cause of the color variability described above,
we utilize the line fluxes inferred from the spectrum shown in Figure
\ref{Steward_spectrum}. We note that the red-shifted CIV line falls
within the B-band range. At $\lambda = 4740$~\AA, the B-band filter
has a transmission coefficient of approximately $f_{\lambda} = 70$~\% 
of its maximum value. Hence, using the bandwidth of the B filter of 
$\Delta \nu = 1.4 \times 10^{14}$~Hz, we find that the CIV line will 
make an effective contribution of $F_{\nu, CIV}^B = F_{CIV} \, 
f_{\lambda} / \Delta\nu \approx 2.1 \, \mu$~Jy to the B-band 
flux. An analogous calculation of the contribution from the 
CIII] line to the V and R bands yields $F_{\nu, CIII]}^V \approx 
1.0 \, \mu$Jy and $F_{\nu, CIII]}^R \approx 1.6 \, \mu$Jy. 
We corrected all B, V, and R magnitude values for
these line contributions and re-evaluated the color magnitude 
correlations. The resulting points are shown as solid symbols
in Figure \ref{color}. This leads to an overall slight shift
towards fainter (larger R) magnitudes and redder B - R colors.
However, the overall color variability trend and its significance
remain unaffected by this correction.

\begin{figure}
\plotone{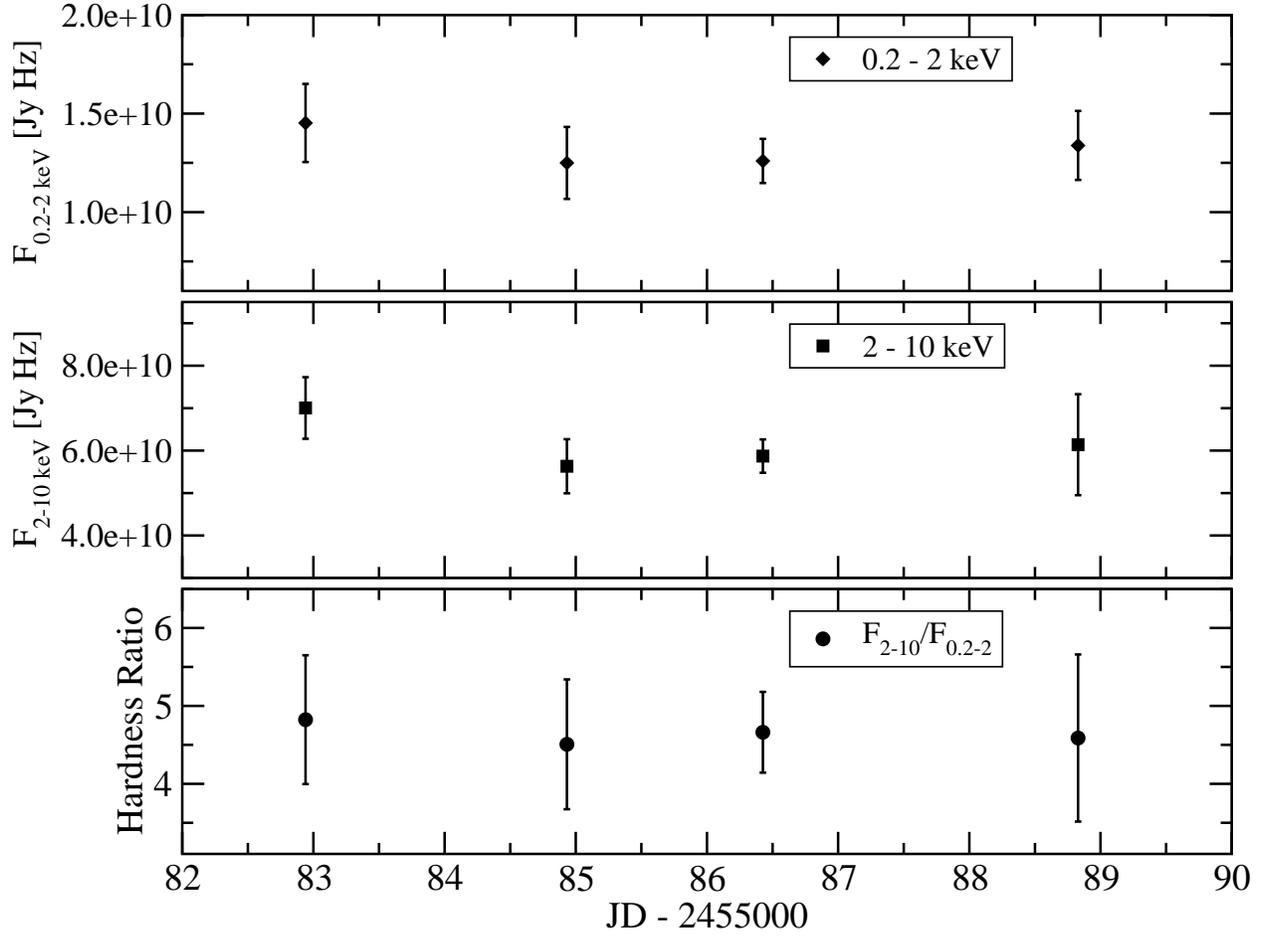}
\caption{Light curves in soft (0.2 -- 2~keV) X-rays (top panel), 
hard (2 -- 10~keV) X-rays (middle), and the hardness ratio. No 
significant spectral variability was found.}
\label{xrate}
\end{figure}

\begin{figure}
\plotone{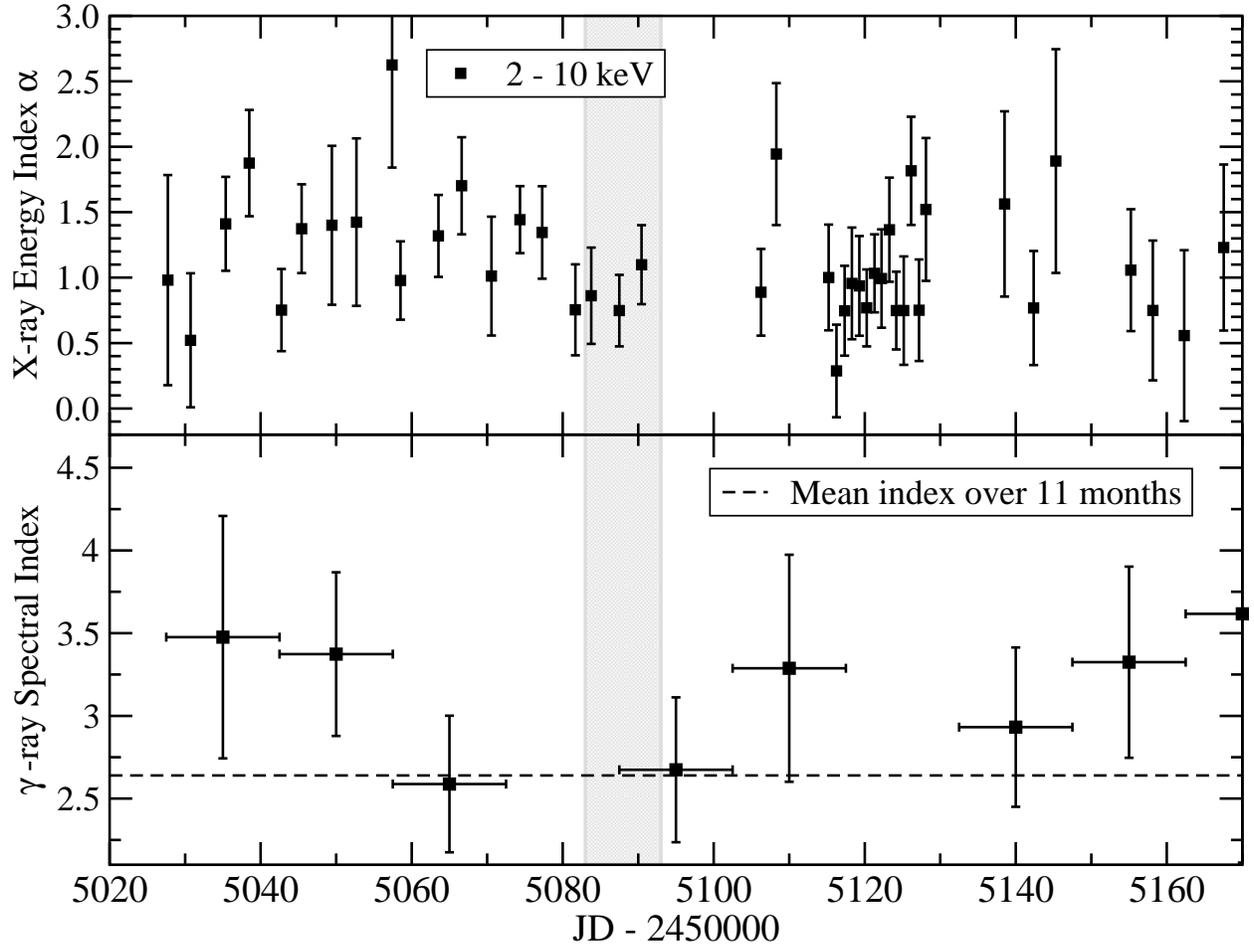}
\caption{Top panel: X-ray (2 -- 10~keV) energy spectral index $\alpha$ vs.
time. Bottom: Fermi $\gamma$-ray photon spectral index ($\Gamma_{\rm ph} = 
\alpha + 1)$ vs. time. The dashed horizontal line corresponds 
to an index of 2.64, the mean index observed over the first eleven months of 
Fermi data. }
\label{xraygamma}
\end{figure}

We also analyzed the spectral variability in the X-ray regime. For the 
XMM-Newton data, we analyzed the variability of the hardness ratio 
$F_{2 - 10 keV}/F_{0.2 - 2 keV}$. Figure \ref{xrate} illustrates that the 
hardness ratio is consistent with being constant over the four 
{\it XMM-Newton} observations performed as part of the core campaign. 
The RXTE X-ray energy index 
$\alpha$ was analyzed over a period of 150 days as plotted in the 
top panel of figure \ref{xraygamma}. A fit to a constant results 
in $\chi_{\nu}^2 = 1.31$ is found, indicating very moderate spectral
variability. The comparison of the {\it XMM-Newton} spectra with the 
{\it Suzaku} spectrum from 2008 September/October (see \S \ref{Suzaku}) 
suggests that the X-ray spectral index remains stable on very long time 
scales, even throughout substantial (factor $\sim 2$ -- 3) flux variations. 

We also analyzed the Fermi $\gamma$-ray spectral index variability over the 
extended campaign period. As shown in the bottom panel of figure \ref{xraygamma}, 
no significant variability is found. The best fit found for the spectral index 
in this period of time is $\Gamma = 2.9 \pm 0.2$, which is slightly softer than 
the mean spectral index ($\Gamma = 2.64 \pm 0.06$) observed over the first eleven 
months of Fermi data \citep[dashed horizontal line][]{abdo10}.

\begin{figure}
\plotone{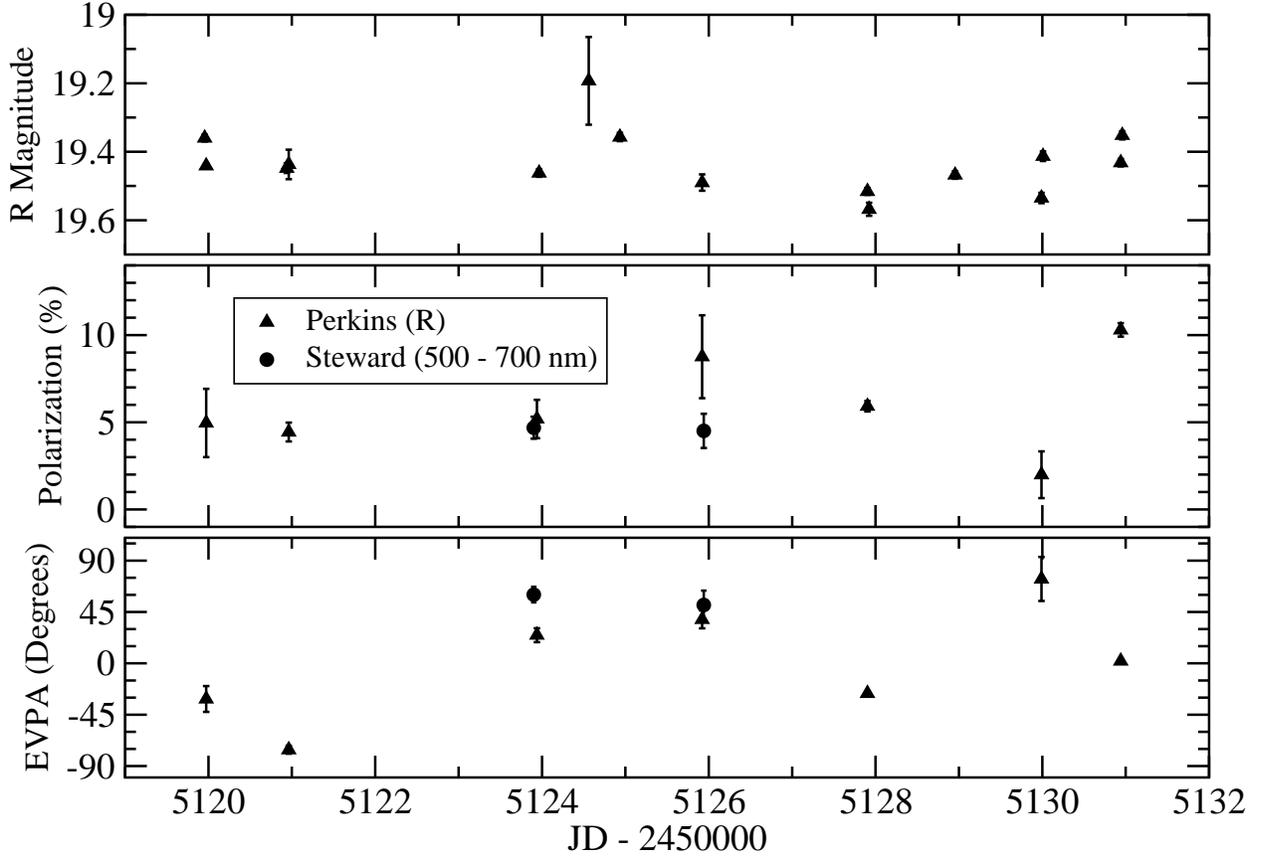}
\caption{Simultaneous photometric and polarimetric data for PKS~0528+134 on
seven days during October 15 -- 26, 2009. Top panel: R-band magnitude. Middle panel: 
Degree of polarization. Bottom panel: Electric vector position angle (EVPA).
While there is significant variability in the R-band flux, degree of polarization,
and EVPA, we did not find a significant correlation between polarization and
flux states. }
\label{polar}
\end{figure}

\subsection{\label{polvar}Optical Polarization Variability Analysis}

The polarization variability analysis on PKS 0528+134 was performed using 
the R-band  polarimetric observations at the 1.8~m Perkins telescope of 
Lowell observatory during a twelve day period from October 15 to 
26, 2009, and two measurements from the University of Arizona / Steward
Observatory {\it Fermi} support program from the same period.
As shown in figure \ref{polar}, 
the Perkins polarimetry data indicate strong variability of 
the degree of polarization ($\chi_{\nu}^2 = 20.2$ for a fit to a constant)
and the electric vector position angle (EVPA). 
The variability of the polarization parameters occurs on time scales of
$\sim 1$ -- 2~days and shows no obvious correlation with the fluxes in 
the optical or any other wavelength band. 

The significant degree of polarization in the optical (and radio)
hints towards a substantial synchrotron contribution to the emission
at these wavelengths. We will
discuss implications of our polarization results in Section \ref{discussion}.

\section{\label{scalejet}Structure of the Parsec Scale Jet}

\begin{figure}
\epsscale{0.35}
\centerline{\includegraphics[width=.32\textwidth]{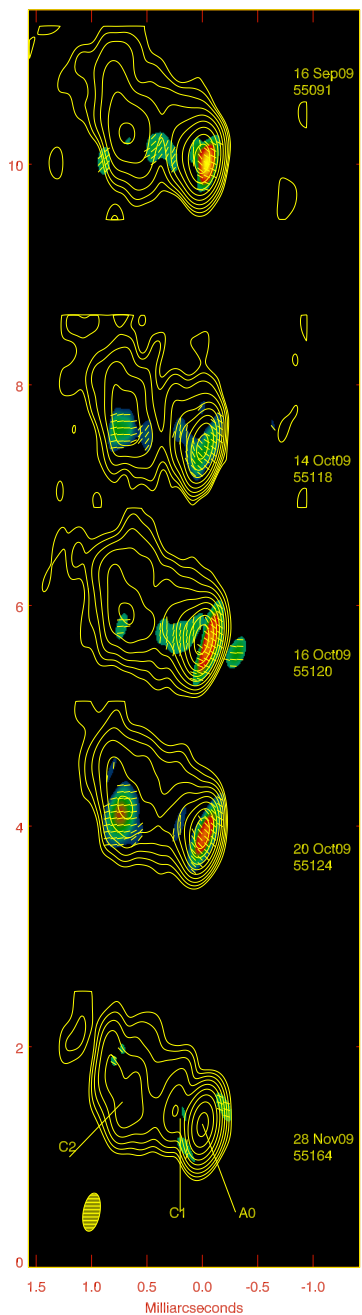}}
\caption{43 GHz total ({\it contours}) and polarized ({\it color scale})
intensity images of PKS~0528+134 during 2009 Autumn. The highest contour
corresponds to $S_{\rm peak}$=850~mJy/beam, while the yellow color indicates
the highest polarized flux of $S_{\rm peak}^{\rm p} = 30$~mJy/beam, for
a beam of $0.33 \times 0.15$~mas$^2$ at PA=-10$^\circ$.
Total intensity contours correspond to 0.25, 0.5, ..., 64 \% of the peak. 
Sticks over the polarized intensity contours indicate the plane of 
polarization. The designation of components corresponds to Table \ref{Knots}.} 
\label{maps43}
\end{figure}

\begin{figure}
\epsscale{1}
\plotone{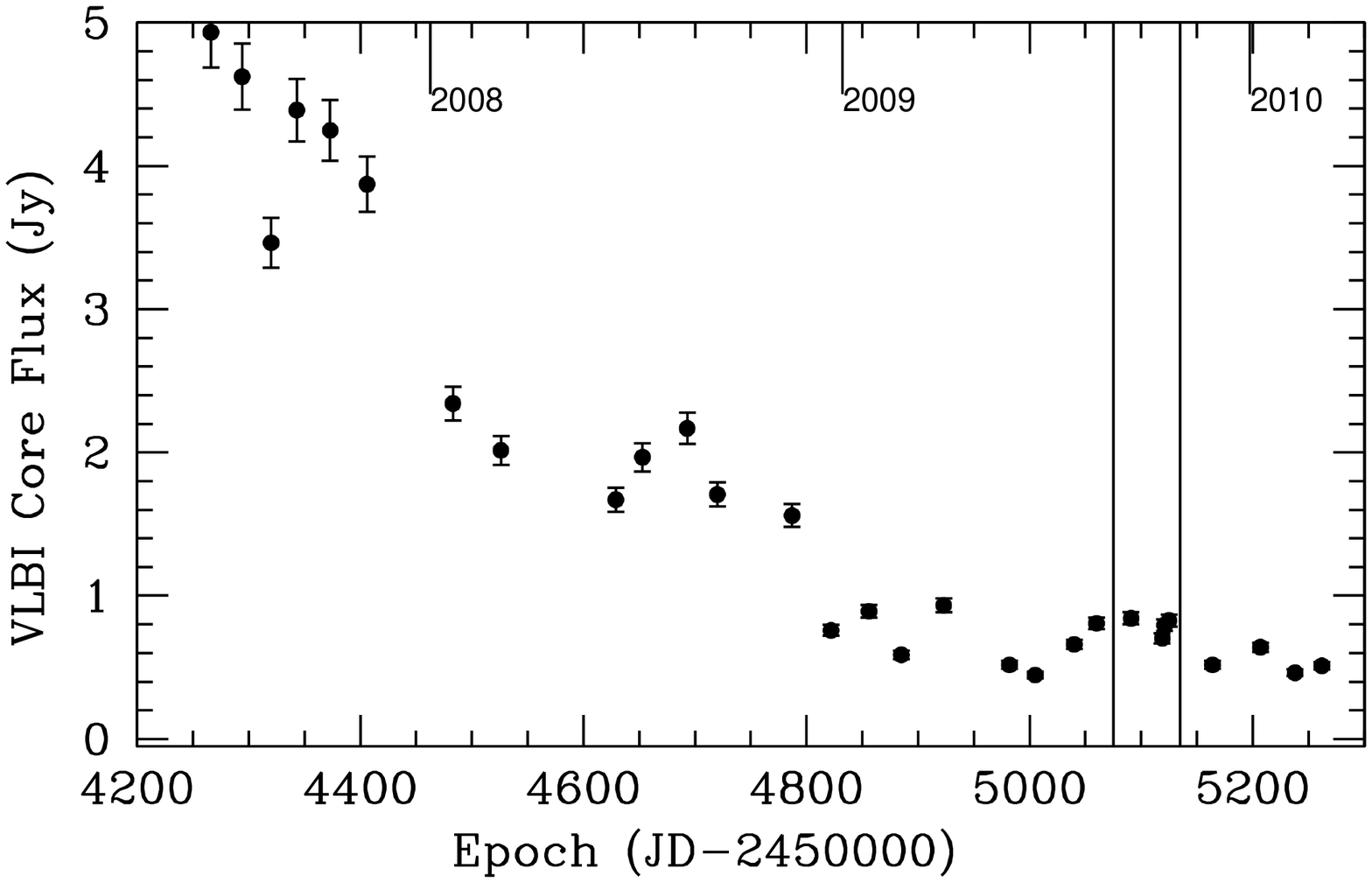}
\caption{Light curve of the VLBI core at 43 GHz. The vertical lines indicate
the period of intensified monitoring in 2009 Autumn. } 
\label{Core}
\end{figure}

The quasar PKS~0528+134 is monitored monthly by the Boston University (BU) group 
with the Very Long Baseline Array (VLBA) at 43~GHz within a sample of bright 
$\gamma$-ray blazars \footnote{http://www.bu.edu/blazars}. The source
was also included in a 2-week campaign of observations of 12 $\gamma$-ray blazars
organized in 2009 October when 3 additional VLBA epochs at 43~GHz
were obtained (VLBA project S2053). Figure \ref{maps43} shows the total and polarized
intensity images of the quasar in 2009 Autumn. The VLBA data were calibrated, imaged,
and modeled in the same manner as discussed in \citet{J05}. 
As we did for the optical data, the polarization parameters were evaluated
taking into account the statistical bias as detailed in \cite{wk74}.
Table \ref{Knots}
gives the parameters (flux, position, size, degree and position angle of polarization)
of the main features seen in the radio jet during this period. Figure \ref{Core} shows
the light curve of the VLBI core at 43 GHz of the quasar
over the last three years as monitored by the BU group. 
According to Figure \ref{Core} the parsesc scale jet of PKS~0528+134 was in
a quiescent state in 2009 Autumn. The core was moderately polarized, $P \sim 2$~\%, 
with a stable position angle of polarization in a range between 42$^\circ$
and 50$^\circ$ that aligns within 10$^\circ$ with the jet direction
as determined by the position angle of the brightest knot $C2$ with respect to 
the core. The closest knot to the core, $C1$, has the highest level of 
polarization, $P \sim 10$~\%, with the position of polarization perpendicular 
to the local jet direction. Although we do not observe superluminal knots in 
2009 Autumn, the quasar is known to have a very high apparent speed in the jet, 
implying a high bulk Lorentz factor, $\Gamma \sim 30$ \citep{J05}.

\begin{deluxetable}{lrrrrrrr}
\singlespace
\tablecolumns{8}
\tablecaption{\bf Parameters of Jet Components \label{Knots}}
\tabletypesize{\footnotesize}
\tablehead{
\colhead{Epoch} & \colhead{Knot} & \colhead{S [Jy]} & \colhead{R [mas]} &
\colhead{$\Theta$ [$^\circ$]} & \colhead{a [mas]} & \colhead{P [\%]} & 
\colhead{$\chi$ [$^\circ$]} \\
\colhead{(1)}&\colhead{(2)}&\colhead{(3)}&\colhead{(4)}&\colhead{(5)}&
\colhead{(6)}&\colhead{(7)}&\colhead{(8)}
}
\startdata
16 Sep&$A0$&0.842$\pm$0.085&0.0&\nodata&0.048&2.5$\pm$0.8&47$\pm$7 \\
&$C1$&0.14$\pm$0.025&0.165$\pm$0.025&71.5$\pm$0.5&0.107&12.3$\pm$2.2&$-$12$\pm$8 \\
&$C2$&0.35$\pm$0.050&0.70$\pm$0.05&64.3$\pm$0.5&0.522&$<$2.0&\nodata \\
14 Oct&$A0$&0.702$\pm$0.065&0.0&\nodata&0.000&1.4$\pm$0.6&49$\pm$8 \\
&$C1$&0.14$\pm$0.025&0.136$\pm$0.025&82.4$\pm$0.5&0.187&8.0$\pm$2.5&0$\pm$10 \\
&$C2$&0.29$\pm$0.050&0.71$\pm$0.05&61.8$\pm$1.0&0.568&5.4$\pm$1.8&93$\pm$5 \\
16 Oct&$A0$&0.794$\pm$0.055&0.0&\nodata&0.035&1.0$\pm$0.5&50$\pm$8 \\
&$C1$&0.25$\pm$0.025&0.127$\pm$0.025&68.7$\pm$0.5&0.250&8.2$\pm$2.0&$-$3$\pm$10 \\
&$C2$&0.37$\pm$0.07&0.72$\pm$0.05&63.4$\pm$0.5&0.510&4.0$\pm$1.5&91$\pm$8 \\
20 Oct&$A0$&0.826$\pm$0.065&0.0&\nodata&0.023&2.1$\pm$0.6&42$\pm$5 \\
&$C1$&0.21$\pm$0.025&0.143$\pm$0.025&69.4$\pm$0.5&0.240&$<$5&\nodata \\
&$C2$&0.36$\pm$0.05&0.72$\pm$0.05&64.2$\pm$0.5&0.490&4.4$\pm$1.5&91$\pm$6 \\
28 Nov&$A0$&0.794$\pm$0.045&0.0&\nodata&0.035&$<$1.0&\nodata \\
&$C1$&0.25$\pm$0.02&0.137$\pm$0.025&68.7$\pm$0.5&0.250&$<$4&\nodata \\
&$C2$&0.37$\pm$0.05&0.72$\pm$0.05&63.4$\pm$0.5&0.510&$<$2.4&\nodata \\
\enddata
\tablecomments{Columns: 1 - epoch of the observation; 2 - component designation; 
3 - flux of component; 
4 -  distance of component from the VLBI core; 5 - position angle of component 
with respect to the core; 
6 - diameter of component; 7 - degree of polarization of component; 8 - position
 angle of polarization of component}
\end{deluxetable}

\section{\label{SED}Spectral Energy Distributions (SEDs) Modeling}

\begin{figure}
\begin{center}
\includegraphics[width=6in]{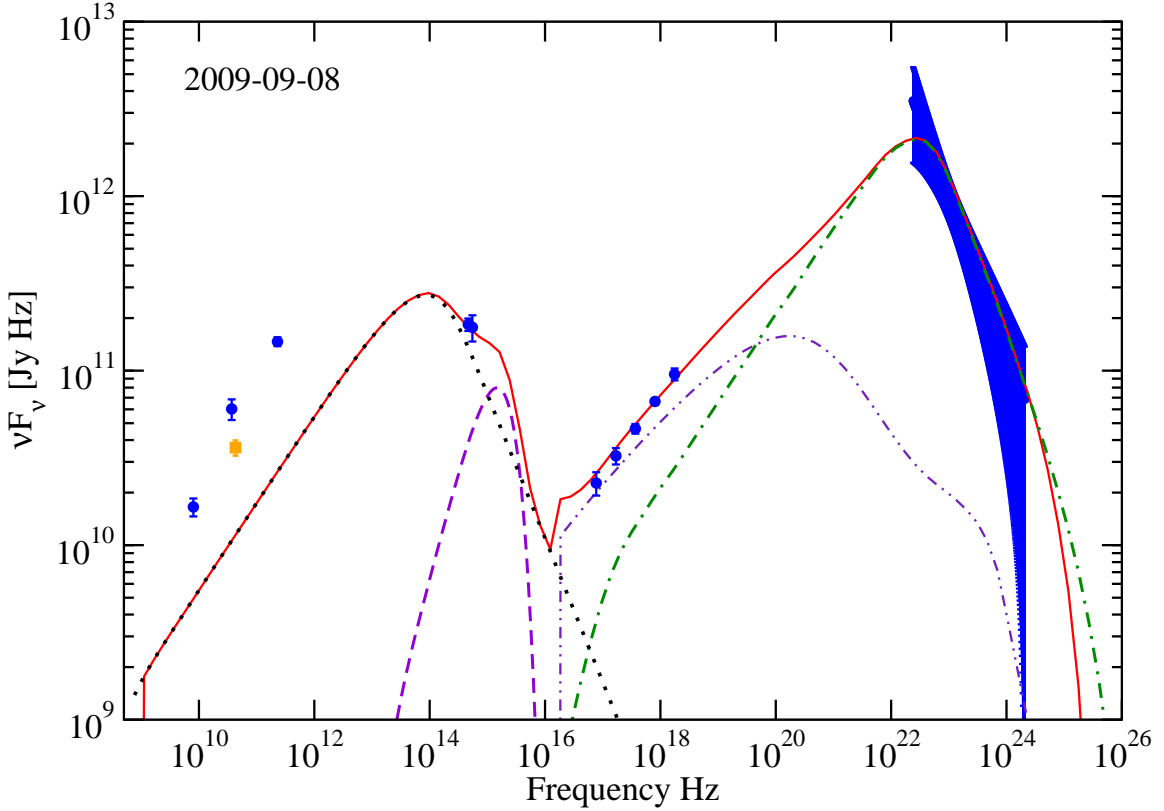} 
\end{center}
\caption{Spectral energy distribution (SEDs) of PKS~0528+134 corresponding 
the first XMM-Newton observation. Blue points correspond to the observational 
data. The radio data correspond to the integrated radio flux from single-dish
measurements. The orange radio point at 43~GHz is the flux of the VLBA core
component.
The continuous lines correspond to  the best fit for each SED. The dotted 
line represents the synchrotron spectrum, the dot-dot-dashed line corresponds 
to the Synchrotron self Compton (SSC), the dot-dashed line is the external 
radiation Compton (ERC), and the short-dashed line is the disk thermal
component.}
\label{seds}
\end{figure}

In figure \ref{seds}, we present the SEDs of 
PKS~0528+134 from radio to  $\gamma$-rays (blue filled circles) corresponding 
to the four XMM-Newton observations (September 8, 10, 11, 14). The optical 
data have been dereddened assuming  $A_V$=2.782,  $A_B = 3.62$ and 
$A_R = 2.24$, as given in the NASA/IPAC Extragalactic 
Database\footnote{http://nedwww.ipac.caltech.edu/} according 
to the sky dust map given by \cite{schlegel98}. We found in sections 
\ref{fluxvar} and \ref{spectvar} that the Fermi $\gamma$-ray flux and spectral 
index of PKS~0528+134 did not show significant variability in the interval 
including the core campaign. Accordingly, the same $\gamma$-ray spectrum
was used in the four SEDs.  As in its flare states, in this quiescent state 
the SED of PKS~0528+134 is characterized by two peaks, a low-energy peak 
between the far infrared and optical spectral bands, and a high-energy peak
at MeV -- GeV energies. As can be seen, in all the SEDs the high-energy 
component dominates the bolometric output by a large amount. 

We produce model fits to all four SEDs using the equilibrium version of 
the leptonic one-zone model developed by \cite{boettcher02}. This equilibrium
model is described in more detail in \cite{acciari09}, and we here summarize
its main features. The observed electromagnetic radiation is interpreted 
as originating from ultrarelativistic electrons (and positrons) in a 
spherical emission region of co-moving radius $R$, which is moving 
with a relativistic speed $\beta_{\Gamma} c$, corresponding to the bulk 
Lorentz factor $\Gamma$. Depending on the viewing angle $\theta$ between 
the jet direction and the line of sight, the transformations of photon 
energies and fluxes is characterized by the Doppler factor $D = \left( 
\Gamma [1 - \beta_{\Gamma} \cos\theta] \right)^{-1}$. The size of the 
emission region is constrained by the shortest observed variability time 
scale $\delta t_{\rm var, min}$ through $R \le c \delta t_{\rm var, min} 
\, D / (1 + z) \lesssim 8 \times 10^{15} (\delta t_{\rm var, min}/{\rm d}) 
\, (D/10)$~cm. 

Ultrarelativistic electrons are assumed to be instantaneously
accelerated at a height $z_0$ above the accretion disk into a 
power-law distribution in electron energy, $E_e = \gamma m_e c^2$, 
at a rate per unit volume and unit Lorentz factor interval given 
by $Q(\gamma) = Q_0 \gamma^{-q}$ with a low- and high-energy cutoffs 
$\gamma_1$ and $\gamma_2$, respectively, and injection spectral index 
$q$. An equilibrium between this particle injection, radiative cooling,
and escape of particles from the emission region yields a temporary
equilibrium state described by a broken power-law. The time scale 
for particle escape is parameterized through an escape time scale 
parameter $\eta_{\rm esc} > 1$ as $t_{\rm esc} = \eta_{\rm esc} R/c$.  
The balance between escape and radiative cooling will lead to a
break in the equilibrium particle distribution at a break Lorentz 
factor $\gamma_b$, where $t_{\rm esc} = t_{\rm cool} (\gamma)$. 
The cooling time scale $t_{\rm cool}$ is evaluated self-consistently
taking into account synchrotron, synchrotron-self-Compton (SSC) 
and external Compton (EC) cooling. The number density of injected 
particles is normalized to the resulting power $L_e$ in 
ultrarelativistic electrons propagating along the jet.
The magnetic field $B$ in the emission region is pre-specified as
a free parameter. It corresponds to a Poynting flux along the
jet, $L_B = \pi R^2 \, \Gamma^2 \beta_{\Gamma} \, c \, u'_B$
where $u'_B = B^2 / (8\pi)$ is the magnetic field energy
density in the co-moving frame.
For each model calculation, the resulting equipartition parameter,
$e_B = L_B/L_e$ is evaluated. 

Once the quasi-equilibrium particle distribution in the emission
region is calculated, our code evaluates the radiative output from 
synchrotron emission, SSC, and EC emission self-consistently with 
the radiative cooling rates. If the occasional indication of a blue 
bump in the optical spectrum can be associated with the accretion 
disk, we can estimate an accretion disk luminosity from the
corresponding approximate $\nu F_{\nu}$ flux in the UV regime of 
$\nu F_{\nu}^{\rm disk} \sim 8 \times 10^{10}$~Jy~Hz as $L_D \sim 1.7 
\times 10^{47}$~erg~s$^{-1}$, which we use for our model fits. 
The accretion disk emission is modelled as a multi-color
blackbody spectrum according to a \cite{ss73} disk model. 

In addition to direct accretion disk emission, external radiation
may originate as line emission from the Broad Line Region (BLR). 
We can estimate the total luminosity of the BLR line emission
using the bright quasar template of \cite{francis91}, normalized
to the observed value of the CIV emission line luminosity of 
$L_{CIV} = 2.5 \times 10^{45}$~erg~s$^{-1}$. Substantial contributions 
to the BLR luminosity, in addition to CIII] and CIV observed here, 
are expected to arise from Fe II, Ly$\alpha$ and Ly$\beta$, 
H$\beta$ and H$\gamma$, Mg II, and He II, among others. The total
BLR luminosity is expected to be $L_{BLR} \approx 7.1 \, L_{CIV}
\approx 1.8 \times 10^{46}$~erg~s$^{-1}$. This is an order of 
magnitude lower than the value we adopt for the accretion disk
luminosity. However, the relevance to external Compton scattering
depends on the photon field energy density in the rest-frame of the
emission region and hence on the geometry and the Doppler boosting 
of external photons into the co-moving frame of the emission region.
The energy density of the direct disk emission in the AGN frame is
$u_{\rm disk} = L_D / (4 \pi z_0^2 \, c)$ where $z_0$ is the distance
of the emission region from the central supermassive black hole. If
$z_0 \gg R_D \, \Gamma^2$, where $R_D$ is the characteristic radius
of the annulus of maximum energy output of the accretion disk, accretion 
disk photons will be strongly red-shifted in the emission-region rest frame. 
However, in the near-field regime, $z_0 \ll R_D \, \Gamma^2$, even the
accretion disk photons entering the emission region from behind in the
AGN rest frame will still be blue-shifted and their energy density 
enhanced in the emission region rest frame \citep[see, e.g.,][]{ds93}.
If the emission region is located within the characteristic radius
of the BLR, $R_{BLR}$, the BLR radiation field can be treated as
approximately isotropic, and it will be blue-shifted into the 
emission-region rest frame, and enhanced by a factor of $\approx 
\Gamma^2$ compared to its AGN rest-frame value of $u_{BLR} = 
L_{BLR} / (4 \pi R_{BLR}^2 \, c)$ \citep{sbr94}.

Lacking knowledge of the size of the BLR and the precise location of
the emission region in PKS~0528+134, both direct disk and BLR emission
contributions remain plausible as dominant sources of external photons
for the external-Compton process. In the following, we have chosen our
model parameters (in particular, $z_0$) such that we expect the direct
disk emission to dominate. As we will see below, this allows for satisfactory
fits to the SED and, in particular, the {\it Fermi}-LAT $\gamma$-ray
emission from PKS~0528+134. Therefore, in order to avoid the introduction
of another unconstrained parameter, $R_{BLR}$, we restrict our modeling 
efforts to using the direct accretion disk emission as the dominant 
contributor to the external radiation field to evaluate the external 
radiation Compton (ERC) component of the SED.

As a representative example, the fit to the SED of PKS~0528+134 
during the first {\it XMM-Newton} observation is shown in Figure \ref{seds}. 
The parameters used for this particular fit are shown in table \ref{param}.
The synchrotron (dotted line), synchrotron self Compton 
(SSC) (dot-dot-dashed line), external radiation Compton (ERC) 
(dot-dashed line), and the direct disk emission (short-dashed line)
components are shown separately, in addition to the total SED fit
curve. 

\begin{deluxetable}{cccccc}
\tabletypesize{\scriptsize}
\tablecaption{Parameters used in the fit for the SED of PKS 0528+134 
corresponding to the XMM-Newton observation of September 8, 2009
(JD~2455082.94).}
\tablewidth{0pt}
\tablehead{
\colhead{Parameter} & \colhead{Value }
}
\startdata
$\gamma_{min}$ &         1400 \\
$\gamma_{max}$ &        $10^5$\\
Injection electron spectral index &	3.65\\
Escape time parameter ($t_{\rm esc} = \eta_{\rm esc} \, R/c$) &	$\eta_{\rm esc} = 50$\\
 Magnetic field  [G] &	  2.05\\
 Injection height [pc] & $z_0 = 0.13$ \\ 
 Bulk Lorentz factor $\Gamma$ & 20.4\\
 Accretion disk luminosity [$10^{46}$ erg/s] & $L_{46} = 17$\\
  Blob radius [cm]  & $1.9 \times 10^{16}$\\
  Black hole mass [$M_{\odot}$]  & $1.08 \times 10^9$ \\
   Observing angle [degrees]  & $\theta_{obs} = 3.02$\\
   Doppler factor & $D = 19.0$\\
Redshift    &                   Z = 2.06\\
 $L_e$ (jet) [erg/s]  & $2.3 \times 10^{45}$ \\
 $L_B$ (jet) [erg/s]  & $2.3 \times 10^{45}$ \\
 $L_B/L_e$              & 1.0\\
 
\noalign{\smallskip\hrule\smallskip}
\enddata
\label{param}
\end{deluxetable}

We were able to achieve good fits for all four SEDs of 
PKS~0528+134 in quiescence. Table \ref{parameters} lists 
the most relevant parameters used in each fit. Fit parameters close to 
equipartition could be found for all four SEDs. In general, no obvious 
correlation between the different parameters was found. However, a strong 
correlation (Pearson's correlation coefficient  $r \approx 1$) was found 
between the magnetic field and the optical flux (R-band), reflecting the
synchrotron dominance in the optical band.

\begin{deluxetable}{ccccccccccc}
\tabletypesize{\scriptsize}
\tablecaption{Relevant fit parameters for the SEDs of PKS~0528+134 }
\tablewidth{0pt}
\tablehead{
\colhead{SED} & \colhead{B [G] } & \colhead{$F_R$ [Jy Hz]} & 
\colhead{$F_X$ [erg cm$^{-2}$ s$^{-1}$]} &\colhead{$\Gamma$} & \colhead{ q}  & 
\colhead{$\eta$ }  & \colhead{$L_e$} & \colhead{$L_B$}  & \colhead{$L_B/L_e$ } & 
\colhead{$z_0$ [pc]}}
\startdata    
1 &   2.05   & 1.8399e+11   & 1.6249e-12  & 20.4  & 3.65  & 50         & 2.30e+45   &  2.30e+45   & 1.0  & 0.13\\
2 &   2.06   & 1.6570e+11   & 1.3224e-12  & 20.1  & 3.85  & 50         & 2.39e+45   &   2.26e+45 & 0.95 & 0.13   \\
3 &	3.7     & 3.8000e+11   & 1.3702e-12  & 17.5  & 3.7     & 70        & 1.18e+45   &   1.73e+45  & 1.46 & 0.11 \\
4 &	2.21   & 1.9799e+11   & 1.4367e-12  & 21.1  & 3.65   & 40.3    &  2.24e+45  &    2.40e+45  & 1.07 & 0.135\\

\noalign{\smallskip\hrule\smallskip}
\enddata
\label{parameters}
\end{deluxetable}

We point out that our model only includes the emission from the
blazar zone, assumed here to be on sub-pc scales. It is expected that
the radio emission originates in the more extended (pc to kpc scales) 
jet which is not included in our model. Therefore, our fits 
under-produce the radio spectra in all SEDs. While most of the
radio data shown in Figure \ref{seds} are obtained by single-dish
instruments and therefore represent the integrated flux over all radio
components, we have also included the 43~GHz flux from the VLBA core
component on 16 September 2009, just 2 days after the last {\it XMM-Newton}
observation. Even this core radio flux is under-represented by our
model, suggesting that the higher-frequency emission originates on
even smaller scales than the radio core.

\section{\label{discussion}Discussion}

As mentioned in the previous section, good fits with a one-zone leptonic 
SSC + ERC jet model were possible with parameters close to equipartition.
Our fit parameters, in particular, the magnetic fields of $B \sim 2$ -- 3~G, 
characteristic electron energies, $\gamma_{\rm min} \sim 10^3$, and jet 
powers of $L_e \sim 10^{45}$~erg~s$^{-1}$, are in rough agreement with the fit
results of \cite{muk99}. However, we need to point out that the model used
in \cite{muk99} is not precisely the same as used in this paper, as it was
based on a time-average of an evolving particle distribution along the jet.
Also, \cite{muk99} used a different cosmology (a matter-dominated Universe 
with $q_0 = 0.5$), resulting in a substantially different luminosity distance. 
Therefore, our results are not directly comparable. We notice that the {\it Fermi}
LAT $\gamma$-ray spectra of PKS~0528+134 are systematically softer than the
photon indices $\Gamma_{\rm ph} \sim 2.2$ -- 2.6 found during the EGRET era
\citep{muk99}. Therefore, our fits to the LAT SEDs require significantly 
steeper particle spectral indices. 

We also need to caution that the model contains a large number of poorly 
constrained parameters, and in many cases, different parameter combinations 
might be able to produce similarly acceptable fits. Therefore, conclusions 
about correlations between model parameters and observables may not be unique. 

A similar study of $\gamma$-ray bright blazars in quiescence has recently been
published by \cite{abdo10b}. Those authors observed 5 blazars (PKS~0208-512,
Q~0827+243, PKS~1127-145, PKS1510-089, and 3C~454.3) in their low-activity
state with Suzaku and {\it Swift} in X-rays and optical/UV, and analyzed the 
simultaneous {\it Fermi}-LAT data. All of those blazars showed X-ray continua
consistent with a hard ($\Gamma_{\rm ph} \sim 1.5$) single power-law spectrum,
in agreement with our {\it XMM-Newton} and {\it Suzaku} results on PKS~0528+134.
The broadband SEDs of all five blazars were dominated by their high-energy
($\gamma$-ray) output, as in PKS~0528+134. In contrast to our {\it XMM-Newton}
results on PKS~0528+134, three of the five blazars observed by \cite{abdo10b}
(PKS~0208-512, PKS~1127-145, and PKS~1510-089) did show significant short-term 
variability on time scales of $\sim 5$ -- 10~hr. However, we need to point out
that the observations by \cite{abdo10b} did not adhere to the strict quiescence
criterion required for triggering our PKS~0528+134 campaign (i.e., the {\it Fermi}-LAT
$\gamma$-ray flux remaining below the lowest EGRET flux or upper limit persistently
for at least two weeks, see \S \ref{intro}). In fact, some of their targets (in
particular, PKS~0208-512, PKS~1510-089, and 3C~454.3) exhibited substantial 
$\gamma$-ray flux and even $\gamma$-ray flares within just a few days of the 
{\it Suzaku} observations. Therefore, the X-ray variability reported in 
\cite{abdo10b} may not be characteristic of a truly quiescent state of those 
blazars as investigated in the case of PKS~0528+134 reported in this paper.

In our analysis of optical spectral variability we found a weak anti-correlation 
between the B-R color and the R-band magnitude. A possible explanation for this 
softer when brighter trend can be found in the interplay between the two radiation 
components that contribute photons to the optical flux: the synchrotron component, 
which is generated in the jet itself, with a steep spectrum, always dominating
at low (R-band) frequencies, and slowly varying emission components associated 
with the accretion disk and the BLR. The direct accretion disk emission is 
expected to peak in the ultraviolet and may therefore contribute at the blue 
end of the optical spectrum. 
In Section \ref{spectvar}, we have shown that the observed color
variability can not be attributed to the contribution from the CIII] and
CIV emission lines in the optical spectrum of PKS~0528+134. Therefore,
the obvious candidate is direct, thermal accretion disk emission. 
The suggestive trend of increasing polarization with increasing
wavelength lends further support to this hypothesis.

Given the evidence we found for a substantial accretion disk with a
luminosity of $L_D \sim 1.7 \times 10^{47}$~erg~s$^{-1}$, it is worth
investigating whether at least part of the observed X-ray emission may
result from Comptonization of soft disk photons in a hot, thermal corona 
above the accretion disk. In the extreme scenario, in which the entire
{\it XMM-Newton} spectrum is produced by the corona, Figure \ref{seds}
illustrates that one would require the energy dissipated in the corona
to be of the same order as that dissipated in the optically thick disk.
In the case of Comptonization of soft photons with energy far below the
X-ray regime, the X-ray spectral index $\Gamma_{\rm ph} \sim 1.6$ can 
then be translated into a Comptonization parameter $y = 16 \, \Theta^2
\, \tau_T \, (1 + \tau_T)$, where $\Theta = k T / (m_e c^2)$ is
the dimensionless coronal temperature, through

\begin{equation}
\Gamma_{\rm ph} = - {1 \over 2} \pm \sqrt{ {9 \over 4} + {4 \over y}}
\label{Gammay}
\end{equation}
yielding a value of $y = 1.85$. This would imply coronal parameters
of $\Theta^2 \, \tau_T \, (1 + \tau_T) = 0.116$. However, in order for
the X-ray spectrum to be dominated by coronal Compton emission, the
jet SSC emission would have to be strongly suppressed compared to the
model fits we presented here. This could, in principle, be achieved by
the choice of a much larger emission region size $R$. Such a choice might
be problematic for two reasons: (1) For $R \gtrsim 1.7 \times 10^{16}
\, (D/20)$~cm, causality would not allow for variability on time scales
of a day or less; (2) as the observed synchrotron emission would require
the same magnetic field as chosen in our SED fits, this would result in
the energy density in the leptonic particle population being far below
equipartition with the magnetic field. While both arguments may not
strictly rule out such a scenario, we strongly prefer our fit scenario,
close to equipartition, and allowing for $\sim$~day scale variability.

We found that the shortest observed variability occurs at optical wavelengths.
In the standard interpretation, this is the region near the high-frequency
end of the synchrotron spectrum, emitted by the highest-energy electrons with
the shortest radiative cooling time scales. The synchrotron cooling time scale
for electrons radiating via synchrotron emission in the V band, is

\begin{equation}
\tau_{\rm sy}^{\rm obs} (V) \approx 2.2 \times 10^4 \, (1 + z)^{1/2} \,
B_G^{-3/2} \, D^{-1/2} \; {\rm s}
\label{tausygeneral}
\end{equation}

Using the redshift of $z = 2.07$, a characteristic magnetic field of
$B \sim 2$~G, and a typical Doppler factor of $D = 19$ from our fit
results, we find an observed synchrotron cooling time of $\tau_{\rm sy}^{\rm obs}
(V) \approx 0.9$~hr. Since the SED of PKS~0528+134 is strongly dominated by
the $\gamma$-ray emission, the actual radiative cooling rate is expected to
be shorter than the rate expected from synchrotron emission alone, by a
factor corresponding to the Compton dominance (the ratio of power output
in the high-energy vs. the synchrotron component), which is of order 10. 
Therefore, the total radiative cooling time scale of electrons emitting
synchrotron in the V band is likely of the order of $\sim 1$~minute or
shorter. This suggests that the light curves of variability in all optical
bands are strongly dominated by light-travel time effects rather than
microphysical (acceleration and cooling) time scales. 

The very moderate degree of variability observed in all wavelength bands,
in tandem with the short radiative cooling time scales, implies on-going
particle acceleration in the quiescent state of PKS~0528+134. On the 
other hand, the significant polarization variability hints towards a 
turbulent process. Steady, on-going particle acceleration can be envisioned
in shear-layers of radially stratified jets or particle acceleration at 
standing features, such as re-collimation shocks. However, the magnetic
fields in shear-layers of radially stratified jets are expected to be
highly ordered due to the directionality of the differential motion in
those systems. Therefore, our results seem to favor scenarios involving
turbulent acceleration, possibly associated with stationary re-collimation
shocks.

Further evidence for turbulent processes comes from our polarization
results. A power-law distribution of non-thermal electrons with a spectral 
index $p$ can produce synchrotron emission with a maximum polarization degree
of $P_s^{\rm max} = (p + 1)/(p + 7/3)$ for a perfectly ordered magnetic
field \citep{rl86}. Based on our modeling results in Section \ref{SED},
the high-energy (cooled) part of the electron spectrum is expected to have
a spectral index of $p = q + 1 \approx 4.6$. This would yield a maximum
polarization degree of $P_s^{\rm max} \approx 70$~\%. The fact that the
actual degree of polarization remains below $\sim 10$~\% indicates the
magnetic field is tangled on size scales much smaller than the size of
the emission region. The rapid (day scale), apparently random variation 
of the degree of polarization and the EVPA seems to indicate a turbulent 
process resulting in a large number of individual cells with randomly 
oriented magnetic fields. Monte Carlo simulations \citep{darcangelo10} 
indicate that for a 100~\% chaotic magnetic field (i.e., no ordered 
field component), 150 turbulence cells result in a degree of polarization 
of $p = (6 \pm 3)$~\%, in agreement with the observed values for PKS~0528+134.
Given a typical variability time scale in the optical of $\sim 1$~day,
the size of individual turbulence cells can be estimated as $R_{\rm cell}
\sim c \, \delta t_{\rm var} / (N^{1/3} \, D) \sim 2 \times 10^{14}$~cm
for a Doppler factor of $D \sim 20$ (see \S \ref{SED}) and $N = 150$ 
turbulence cells. Comparing the optical polarization variability to the
polarization variability of the radio core (see \S \ref{scalejet}), one
can speculate that when the optical polarization is high ($\gtrsim 5$~\%),
the optical EVPA seems to be more closely aligned with the EVPA of the
radio component closest to the radio core.

\section{\label{summary}Summary and Conclusions}

Over the last two decades PKS~0528+134 has become an important target for 
multiwavelength observations because of its high luminosity from radio 
through $\gamma$-rays and the extreme flux and spectral variability that 
it shows in its flaring states. In this paper, we have presented multiwavelength 
observations of PKS~0528+134 involving the {\it XMM-Newton}, RXTE, {\it Suzaku},
and {\it Fermi} satellites as well as many ground based radio and optical 
telescopes. Our main goal was to characterize this $\gamma$-ray loud quasar 
in a quiescent state to improve the understanding of SEDs and variability 
patterns of this source, and of blazars in general. The variability analysis 
of the collected data, and the construction and modeling of four SEDs of this 
source, yielded the following results:

\begin{itemize}

\item No significant short term flux and spectral variability (as determined by 
data in the core campaign) was found in $\gamma$-rays, X-rays and most radio 
bands. However, for the same time interval, significant flux variability 
with $\Delta R \lesssim 1^{\rm mag}$
on time scales of several hours was found in the optical, accompanied by 
a weak spectral softening with increasing flux. The latter trend may be 
interpreted as a steady contribution of the accretion disk flux 
at the blue end of the optical spectrum. 
Optical spectropolarimetry suggests an increasing degree of polarization
towards longer wavelengths, lending further support to the hypothesis of 
synchrotron emission contributing an increasing fraction of the emission
towards the red end of the spectrum.

\item Data analysis based on a more extended interval (two months or more) shows 
no significant $\gamma$-ray flux and spectral index variations, but moderate 
flux variability in the X-rays ($\vert\Delta F / F \vert \sim 50$~\%)
and at radio ($\vert\Delta F / F \vert \lesssim 20$~\%)
frequencies on time scales of
$\sim 1$ -- 2~weeks. A comparison between {\it XMM-Newton} spectra of
2009 September and the {\it Suzaku} spectrum of 2008 September/October
suggests that the X-ray spectral index remains stable in the quiescent
state of PKS~0528+134 even throughout substantial (factor $\sim 2$ -- 3)
long-term flux variations.

\item We constructed four SEDs of PKS~0528+134 in quiescence. Our results show 
that even in the quiescent state, the bolometric luminosity of PKS~0528+134 
is strongly dominated by its $\gamma$-ray emission, although the $\gamma$-ray 
spectra are significantly steeper than found during the EGRET era. 

\item We fitted the four SEDs with a leptonic combined SSC+ERC jet model. In
this model, the low energy component is produced by the sum of the synchrotron 
process in the jet and the disk luminosity, and the high energy emission 
is due to the sum of the synchrotron self-Compton and the external radiation 
Compton contributions. Fit parameters close to equipartition were found for 
all SEDs.

\item The moderate variability in most wavelength bands, compared to the 
expected short radiative cooling time scale, implies the persistence
of particle acceleration on long time scales. This may favor acceleration
scenarios based on standing features, such as re-collimation shocks. 

\end{itemize}

\acknowledgments
This work was supported by NASA through XMM-Newton Guest Observer Program 
awards NNX08AD67G and NNX09AV45G, Chandra Guest Observer Program award 
GO8-9100X, and Fermi Guest Investigator Program award NNX09AT82G.

Norman I. Palma Cruz thanks the Fulbright Program and the National 
Autonomous University of Honduras for making his stay in the U. S. 
possible during the interval that this research took place.

The \textit{Fermi} LAT Collaboration acknowledges generous ongoing support
from a number of agencies and institutes that have supported both the
development and the operation of the LAT as well as scientific data analysis.
These include the National Aeronautics and Space Administration and the
Department of Energy in the United States, the Commissariat \`a l'Energie Atomique
and the Centre National de la Recherche Scientifique / Institut National de Physique
Nucl\'eaire et de Physique des Particules in France, the Agenzia Spaziale Italiana
and the Istituto Nazionale di Fisica Nucleare in Italy, the Ministry of Education,
Culture, Sports, Science and Technology (MEXT), High Energy Accelerator Research
Organization (KEK) and Japan Aerospace Exploration Agency (JAXA) in Japan, and
the K.~A.~Wallenberg Foundation, the Swedish Research Council and the
Swedish National Space Board in Sweden.

The VLBA is an instrument of the National Radio Astronomy Observatory, a
facility of the NSF, operated under cooperative agreement by Associated 
Universities, Inc.

Additional support for science analysis during the operations phase is gratefully
acknowledged from the Istituto Nazionale di Astrofisica in Italy and the Centre 
National d'\'Etudes Spatiales in France.

This paper is partly based on observations carried out at the German-Spanish 
Calar Alto Observatory, which is jointly operated by the MPIA and the IAA-CSIC,
and on observations with the Medicina and Noto telescopes operated by INAF --- 
Istituto di Radioastronomia.   

Calar Alto data were acquired as part of the MAPCAT (Monitoring AGN with
Polarimetry at the Calar Alto Telescopes) project.

Acquisition of the MAPCAT data is supported in part by MICIIN (Spain) grants 
AYA2007-67267-C03-03 and AYA2010-14844, and by CEIC (Andaluc\'{i}a) grant
P09-FQM-4784.
 
"The Submillimeter Array is a joint project between the Smithsonian
Astrophysical Observatory and the Academia Sinica Institute of Astronomy
and Astrophysics, and is funded by the Smithsonian Institution and the
Academia Sinica."
 
The Mets\"ahovi team acknowledges the support from the Academy of Finland
to our observing projects (numbers 212656, 210338, and others) 

The research at Boston University (BU) was funded in part by NASA Fermi
Guest Investigator grants NNX09AT99G and NNX08AV65G, and by the NSF
through grant AST-0907893. The PRISM camera at Lowell Observatory was 
developed by K.\ Janes et al. at BU and Lowell Observatory, with funding 
from the NSF, BU, and Lowell Observatory.

D.M. acknowledges support from Russian RFBR foundation via grant 09-02-00092.

The research at UMRAO was funded in part by NSF grant AST-0607523 and  
by NASA grants NNX09AU16G and NNX10AP16G.
The operation of UMRAO is made possible by funds from  the University  
of Michigan.

G. Madejski acknowledges support from NASA through {\it Suzaku} Guest
Observer grant NNX08AZ89G.

The Steward Observatory {\it Fermi} Support Program is supported by NASA
through {\it Fermi} Guest Investigator Program grants NNX08AW56G and
NNX09AU10G.

\end{document}